\documentclass[twocolumn]{aastex631}
\usepackage{amsmath,amssymb,amsthm,amsbsy,bm}
\usepackage{wasysym}
\usepackage{graphicx,subfigure,tabularx}
\usepackage{natbib}
\usepackage{CJK}
\usepackage{enumitem}
\usepackage{cancel}
\usepackage{color}
\usepackage{ulem}
\usepackage[mathlines]{lineno}
\usepackage{comment}
\hypersetup{linktocpage}

\begin{document}
\begin{CJK*}{UTF8}{gbsn}

\title{Magnetic field of gas giant exoplanets and its influence on the retention of their exomoons}

\author{Xing Wei (魏星)}
\affil{Institute for Frontier in Astronomy and Astrophysics, Department of Astronomy, Beijing Normal University, Beijing, China}

\author{D.N.C. Lin (林潮)}
\affil{Department of Astronomy and Astrophysics, UCSC, Santa Cruz, USA}
\affil{Institute for Advanced Studies, Tsinghua University, Beijing, China}

%\author{Andrew Cumming}
%\affil{Department of Physics and McGill Space Institute, McGill University, Montreal, Canada}

\correspondingauthor{Xing Wei}
\email{xingwei@bnu.edu.cn}

\begin{abstract}
We study the magnetic and tidal interactions of a gas-giant exoplanet with its host star and with its exomoons, and focus on their retention. We briefly revisit the scaling law for planetary dynamo in terms of its mass, radius and luminosity.
Based on the virial theorem, we construct an evolution law for planetary magnetic field and find that its initial entropy is important for the field evolution of a high-mass planet. We estimate the magnetic torques on orbit arising from the star-planet and planet-moon magnetic interactions, and find that it can compensate tidal torques and bypass frequency valleys where dynamical-tide response is ineffective. For exomoon's retention we consider two situations. In the presence of a circumplanetary disk (CPD), by comparison between CPD's inner and outer radii, we find that planets with too strong magnetic fields or too small distance from its host star tend not to host exomoons. During the subsequent CPD-free evolution, we find, by comparison between planet's spindown and moon's migration timescales, that hot Jupiters with periods of several days are unlikely to retain large exomoons, albeit they could be surrounded by rings from the debris of tidally disrupted moons. In contrast, moons, if formed around warm or cold Jupiters, can be preserved. Finally, we estimate the radio power and flux density due to the star-planet and planet-moon magnetic interactions and give the upper limit of detection distance by FAST.
\end{abstract}

\section{Introduction}

Planetary or stellar magnetic fields influence the structure of surrounding circumstellar and circumplanetary disks during their infancy, their rates of mass and angular momentum loss throughout their life span, and their dynamical interaction with their companions.  These fields are generated by the dynamo action in their convective regions. Thermal convection drives the motion of conducting fluid which interacts with the magnetic field lines to induce new lines to offset magnetic diffusion, i.e. the dynamo action. The dynamo model was initially proposed by \citet{larmor} to interpret the solar magnetic fields.  This model was later applied to geomagnetism by \citet{bullard}. The dynamo theory, especially how turbulent motion generates magnetic fields, was discussed in detail by \citet{moffatt} and \citet{krause}. These theoretical constructs were followed by several numerical simulations of kinematic dynamos \citep{gubbins, galloway, wei2012, wei2014}.  In these kinematic dynamo simulations, the growth of magnetic field is computed for a given (assumed) fluid velocity in the absence of the back reaction of magnetic field on fluid motion. The first three-dimensional (3D) self-sustained dynamos were numerically simulated by \citet{glatzmaier}. This simulation includes the magnetic feedback on the flow.  Subsequently, several follow-up 3D numerical dynamo simulations were carried out \citep{jones2002, christensen2006, wei2018dynamo}.

Due to the enormous computational challenges in the range of spacial and temporal scales, it is not possible to carry out simulations with appropriate magnitude for the relevant physical quantities.  In all the simulations, some gross approximations have to be introduced.  They are usually carried out with some manageable, but unrealistic, adopted model parameters. For example, the real Ekman number (ratio of rotational timescale to viscous timescale) in the Earth's core is at the order of $10^{-15}$.  But, the smallest Ekman number that can be achieved in the most recent geodynamo simulations is at the order of $10^{-7}$. For Jovian dynamo, in addition to Jupiter's fast rotation and small viscosity, the electric conductivity in its interior varies over orders of magnitude.  All of these properties add to the technical difficulties for the numerical studies of planetary dynamo. In order to link the results obtained with artificial numerical models to actual planetary contexts, it is necessary to find a scaling law \citep{christensen2006, jones2014}. Such extrapolation has been applied to the planets in the present-day solar system \citep{christensen2009, davidson2013} and the evolution of exoplanets \citep{reiners2010}. A summary of some scaling laws can be found in \citet{christensen2010}. On the other hand, the internal structure of planets, especially gas giant planets, has evolved considerably since the epoch of their formation, and the initial entropy is found to have an impact on the structure evolution of a high-mass planet \citep{marley2007}, which might influence the field evolution. In this paper we derive a scaling law and an evolution law of planetary magnetic field in a more rigorous way and then apply it to young planets. 

The star-planet or planet-moon magnetic interaction plays an important role in the orbital dynamics, especially at the young age when stellar and planetary fields are strong \citep{zarka2007, laine2008, strugarek2017}. With the estimated magnetic fields, we can interpret moon's 
orbital migration in a circumplanetary disk and find the possibility for the retention of exomoons around the different host planets. We can also estimate the radio emissions due to the star-planet or planet-moon magnetic interaction.

The paper is structured as follows. In \S\ref{sec:field} we derive the scaling law for planetary dynamo (\S\ref{sec:scalinglaw}), find the evolution law and estimate the magnetic field of young gas giant planets with different mass(\S\ref{sec:evolution}). In \S\ref{sec:magnetic-torque} we discuss the various magnetic interactions in a star-planet or planet-moon system (\S\ref{sec:magnetic-interaction}), summarize all the magnetic and tidal torques (\S\ref{sec:all-torques}), and compare their relative strength (\S\ref{sec:magtide}). In \S\ref{sec:moon} we apply the tidal and magnetic torques to the retention of exomoons in a close system by comparison between the inner and outer radii of circumplanetary disk (\S\ref{sec:radii}) and between planet's spindown and moon's migration timescales (\S\ref{sec:migration}). In \S\ref{sec:radio} we estimate the radio power and flux density due to star-planet and planet-moon magnetic interactions and give the upper limit of detection distance by FAST. In \S\ref{sec:discussions} some discussions are given.

\section{Magnetic field of a gas giant planet}\label{sec:field}

In this section we discuss the planetary dynamo and planetary field evolution.

\subsection{Planetary field dynamo}\label{sec:scalinglaw}

In this subsection we estimate the planetary magnetic field in terms of mass, radius and luminosity. Here we briefly summarize the results and the details can be found in Appendix A. In the first place, we derive the total energy equation to find the power balance, i.e., buoyancy power $\sim$ ohmic dissipation rate. Next by the two-scale analysis on the length scale of magnetic field, we derive the energy equipartition, i.e., magnetic energy $\sim$ buoyancy energy. According to the standard mixing length theory, buoyancy energy $\sim$ kinetic energy. Eventually, we estimate the convective velocity and mixing length by the Lane-Emden equation and find the magnetic energy as a function of mass, radius and luminosity
\begin{equation}\label{estimation}
\langle B^2/\mu\rangle \approx 0.115 M^{1/3}R^{-7/3}L^{2/3}.
\end{equation}

Magnetic energy arises from thermal convection on short timescale and thermal convection arises from gravitational contraction on long timescale. Gravitational contraction leads to luminosity (see the next section). Equation \eqref{estimation} reveals the relation between magnetic energy and luminosity. This estimation yields Jupiter's volume-averaged field $\langle B \rangle\approx 70$ gauss, where Jupiter's intrinsic luminosity is $4.60\times 10^{24}$ erg/s \citep{li2018}. The dynamo simulations suggest that $\langle B\rangle \approx 7 B_{dip}$ where $B_{dip}$ is dipole field at surface of dynamo region \citep{christensen2009}. Therefore, our estimation is in good agreement with Jupiter's present dipole field $\approx 10$ gauss. The estimation \eqref{estimation} is what we will use in the next section.

\subsection{Planetary field evolution}\label{sec:evolution}

In this subsection we study the evolution of magnetic field of gas giant planets with different mass. We apply the virial theorem to the evolution of gas giant planet. For a polytropic gas, the relation of gravitational potential energy and internal energy is $E_g=-(3/n)E_i$ \citep{chandrasekhar1939}.
With a non-uniform density profile, the gravitational potential energy reads $E_g=-3GM^2/(5-n)R$. 

In a planet in the absence of nuclear reaction, its luminosity is the rate of total energy $L=-d(E_g+E_i)/dt=-(3-n)/(5-n)(GM^2)/R^2(dR/dt)$ where the prefactor $(3-n)/(5-n)$ is $0.43\sim 0.5$ for $n=1\sim 1.5$. Then we simply assume that the evolution of planet radius follows a power law $R\propto t^{-\gamma}$, and inserting it into the luminosity expression we readily obtain $L \propto M^2 t^{\gamma-1}$. Usually $\gamma$ is a small number, e.g., for 1 $M_J$ $\gamma\approx 0.044$ fitted from the numerical results in \citet{marley2007}, and thus $L\propto M^2t^{-1}$ which is in good agreement with the numerical simulations \citep{burrows1997}. 

Inserting the evolution of radius and of luminosity into the estimation \eqref{estimation}, we obtain the evolution of magnetic field
\begin{equation}\label{evolution-b}
\langle B \rangle \propto M^{5/6} t^{(3/2)\gamma-1/3}\propto M^{5/6} t^{-0.267}
\end{equation}
where $\gamma\approx 0.044$ of a low-mass planet is used. 
Nowadays Jupiter's surface dipole field at 4.6 Gyr is about 10 gauss at poles. We then use the evolution law \eqref{evolution-b} to extrapolate Jupiter's surface dipole field at the two young ages, one at 2 Myr when the protoplanetary disk has a significant impact on dynamics and the other at 10 Myr when the protoplanetary disk almost disappears. The results are $B_{dip}\approx 80$ gauss at 2 Myr and $B_{dip}\approx 50$ gauss at 10 Myr.

Our evolution law shows that magnetic field depends on mass. Moreover, different mass leads to different tracks of planetary evolution \citep{marley2007,fortney2009}, and consequently different field strength. On the other hand, although the initial condition for planetary evolution is eventually forgotten, the planetary evolution at young age greatly depends on the initial condition, namely whether a hot start with a high entropy or a cold start with a low entropy due to the different core-accretion formation models (in the former heat release by accretion shock is transferred to planet interior whereas in the latter heat is radiated away). 

For 1 $M_J$ the two initial conditions lead to almost the same evolution tracks. But for higher mass the evolution tracks with a hot versus cold start differ more. Therefore, the dipole field and dipole moment of young planets are primarily determined by mass and the initial condition for evolution. Since the initial entropy does not influence the evolution track of a low-mass planet, we use 1 $M_J$ as reference to estimate magnetic field of 1 $M_J$ at different age, and then use \eqref{estimation} with the numerical results in \citet{marley2007} to estimate magnetic field of high-mass planets.

Table \ref{table1} shows the magnetic field of planets with 1 $M_J$, 4 $M_J$ and 10 $M_J$ at 2 Myr, 10 Myr and 5 Gyr for the two initial conditions (we do not need to list the two initial conditions for 1 $M_J$ because they yield the same results). Clearly, {\it a larger mass and the hot start correspond to stronger magnetic field}. At 2 Myr, a 10 $M_J$ planet has a magnetic field as high as $\sim$400 gauss with the hot start. But with the cold start, it is $\sim$100 gauss. Moreover, {\it magnetic field decays much sharper with the hot start than with the cold start}. \citet{reiners2010} uses the evolution model with the hot start results in \citep{burrows1997} to estimate dipole field and their results are comparable to ours.
\begin{table*}
\centering
\begin{tabular*}{0.55\textwidth}{@{\extracolsep{\fill}}|c|c|c|c|c|c|}
\hline
       & 1 $M_J$ & 4 $M_J$ cold & 4 $M_J$ hot & 10 $M_J$ cold & 10 $M_J$ hot \\ \hline
2 Myr  & 79      & 102          & 273         & 118           & 438          \\ \hline
10 Myr & 51      & 87           & 155         & 110           & 342          \\ \hline
5 Gyr  & 10      & 13           & 13          & 26            & 26           \\ \hline
\end{tabular*}
\caption{Magnetic field (gauss) for different mass and age with different initial condition.}\label{table1}
\end{table*}

\section{Tidal and magnetic interactions in star-planet and planet-moon systems}\label{sec:magnetic-torque}

In this section we discuss the various magnetic interactions and torques, and then list all the tidal and magnetic torques in the star-planet and planet-moon systems.

\subsection{Star-planet or planet-moon magnetic interactions}\label{sec:magnetic-interaction}

In this subsection we estimate the magnetic torque on orbit arising from star-planet or planet-moon magnetic interaction. The star-planet magnetic interaction consists of three types \citep{zarka2007}: the interaction of unmagnetized stellar wind and planetary field, the interaction of stellar dipole field and planetary dipole field (i.e., the dipole-dipole interaction) \citep{strugarek2015,strugarek2016,strugarek2017} and the interaction of stellar field and orbital motion of unmagnetized planet (i.e., the unipolar induction) \citep{goldreich1969,laine2008,laine2012}. In addition to these three types, the interaction of the misaligned field and rotation of host star also leads to the time-dependent magnetic interaction with its planet \citep{laine2008}, which is relatively weak and will not be studied in this paper. The planet-moon magnetic interaction consists of two types: the dipole-dipole interaction (e.g., Jupiter-Ganymede) and the unipolar interaction (e.g., Jupiter-Io).

\subsubsection{Star-planet dipole-dipole interaction}
\label{sec:starplanetmag}
We now consider the star-planet magnetic torque for the first two interactions, namely the interaction of unmagnetized stellar wind and planetary field and the dipole-dipole interaction. We use a unified empirical formula for the both magnetic torques \citep{strugarek2015}, $\Gamma\approx C_d \cdot a \cdot A_{\rm obs} \cdot P_t$, where $C_d$ is the drag coefficient, $a$ is the orbital radius, $A_{\rm obs}=\pi R_{\rm obs}^2$ is the obstacle area, and $P_t$ is the stellar total pressure at the obstacle radius $R_{\rm obs}$. The drag coefficient is expressed as $C_d\approx M_a/\sqrt{1+M_a^2}$ \citep{zarka2007}, where $M_a$ is the Alfv\'en Mach number $M_a=v/v_a$. Here $v$ is the relative velocity $v=|\omega_p-\Omega_\star|a$ where $\omega_p$ is planet's orbital frequency and $\Omega_\star$ star's rotation frequency, and $v_a=B_\star(R_\star/a)^3/\sqrt{\mu m_pN}$ is the Alfv\'en velocity where $N$ is the plasma number density, $m_p$ the proton mass and $\mu$ the magnetic permeability. 

For an illustrative estimate, we take the typical values for a close-in star-planet system, i.e., a=0.05 au, $M_\star=M_\odot$ and the rotation period ($2 \pi/\Omega_\star$) of a young star about 3 days. We then readily obtain $v=|\omega_p-\Omega_\star|a\approx 50$ km/s. We take stellar dipole field $B_\star\approx 1500$ gauss \citep{t-tauri2017} and $R_\star\approx 2R_\odot$. The mass loss rate ${\dot M}_\star = 4\pi a^2Nm_pv$ evolves roughly as $t^{-2}$ \citep{wood2002,wood2005}, e.g., the solar mass loss rate at present is $\sim 10^{12}~{\rm g/s}$. We assume that the wind velocity keeps about 400 km/s (this assumption is valid in Parker's model) so that density also evolves as $t^{-2}$ and scales with distance as $a^{-2}$. The density can be estimated as
\begin{equation}\label{density}
N=5(t/4.6{\rm Gyr})^{-2}(a_p/1{\rm au})^{-2}{\rm cm^{-3}}.
\end{equation}
At present $N\approx 5 \mbox{ cm}^{-3}$ at 1 au, and density can be estimated from ${\dot M}_\star$ scaled to an young age, say, 5 Myr, $N\approx 1.7\times10^9 \mbox{ cm}^{-3}$ at 0.05 au. As a result $v_a\approx 500$ km/s, $M_a\approx 0.1$ and $C_d\approx 0.1$. 

The obstacle radius is determined by the balance between $P_t$ and the planetary magnetic pressure, $P_t\approx\left[B_p(R_p/R_{\rm obs})^3\right]^2/\mu$. $P_t$ is the sum of the stellar-wind thermal pressure, the stellar-wind kinetic pressure $P_k=Nm_pv^2$ and the stellar magnetic pressure $P_m=\left[B_\star(R_\star/a)^3\right]^2/\mu$. In the first interaction, namely the interaction of unmagnetized stellar wind and planetary field, $P_m\approx 0$, such that we choose $P_t\approx P_k$. In the second interaction, namely the dipole-dipole interaction, the numerical result \citep{strugarek2016} shows magnetic pressure dominates over kinetic pressure, such that we choose $P_t\approx P_m$. At young age of a star-planet system, stellar field $\approx 1500$ gauss and planetary field $\approx 100$ gauss as estimated in the last subsection, and thus $P_m\gg P_k$, namely the dipole-dipole interaction dominates over stellar-wind-planet-field interaction. For the dipole-dipole interaction, by $\left[B_\star(R_\star/a)^3\right]^2/\mu\approx\left[B_p(R_p/R_{\rm obs})^3\right]^2/\mu$ to derive $R_{\rm obs} \approx R_p(B_p/B_\star)^{1/3}(a/R_\star)$. We insert the typical values: $a=0.05$ au, $R_\star=2R_\odot$, $B_\star\approx 1500$ gauss and $B_p\approx 100$ gauss, and then obtain the obstacle radius $R_{\rm obs}\approx 2.2R_p$. Consequently, we derive the magnetic torque due to the dipole-dipole interaction
\begin{equation}\label{dipole-dipole-torque}
\begin{aligned}
\Gamma^m _{\star p} &\approx \pi C_d(B_\star^2/\mu)aR_{\rm obs}^2(R_\star/a)^6 \\ 
&\approx \pi C_d(B_\star^2/\mu)aR_p^2(B_p/B_\star)^{2/3}(R_\star/a)^4. 
\end{aligned}
\end{equation}
Here the subscript $\star$ denotes star (primary) and $p$ planet (secondary), and the superscript $m$ magnetic torque.

\subsubsection{Star-planet unipolar interaction}
\label{sec:starplanetunip}
We next consider the star-planet magnetic torque in the third interaction, namely the unipolar induction. The electric field induced by the planet's orbital motion on the stellar field is $E=vB_\star(R_\star/a)^3$ where $v$ is the relative velocity as before, the voltage imposed on the planet is $V=E \cdot 2R_{\rm obs}=2vB_\star(R_\star/a)^3R_{\rm obs}$, and then the electric current is $I=2vB_\star(R_\star/a)^3R_{\rm obs}/\Lambda$ where $\Lambda$ is the resistance. The Lorentz force can be calculated as $I \cdot B_\star(R_\star/a)^3\cdot 2R_{\rm obs}$, and by $R_{\rm obs}\approx R_p(B_p/B_\star)^{1/3}(a/R_\star)$ the magnetic torque due to unipolar induction reads
\begin{equation}\label{unipolar-torque}
\begin{aligned}
\Gamma^m _{\star p} & \approx 4(vB_\star^2/\Lambda)aR_{\rm obs}^2(R_\star/a)^6 \\
&\approx 4(vB_\star^2/\Lambda)aR_p^2(B_p/B_\star)^{2/3}(R_\star/a)^4.
\end{aligned}
\end{equation}

The torque $\Gamma^m_{\star p}$ in both Eqs (\ref{dipole-dipole-torque}) and (\ref{unipolar-torque}) is in the direction of 
the planet's motion relative to the star.
In the case that the stellar rotation is in the same direction as the orbital motion, outside the corotation radius magnetic torque induces an outward orbital migration, whereas inside the corotation radius it induces an inward migration. In the case that the stellar rotation and orbital motion are opposite to each other, no matter what the direction of relative velocity is, magnetic torque always retards orbital motion and induces an inward migration. Consequently, magnetic torque and tidal torque are always in the same direction with respect to orbital migration.

In the star-planet unipolar interaction, the resistance is higher at the field footpoint of planetary plasma envelope than planetary interior \citep{laine2012}. According to the Alfv\'en wing model \citep{Neubauer1980,zarka2007,strugarek2016}, the resistance of Alfv\'en wing at magnetosphere reads $\Lambda_1\approx\mu v\sqrt{1+M_a^2}/M_a\approx\mu v/C_d$ where the expression of $C_d=M_a/\sqrt{1+M_a^2}$ is employed (de Colle et al in preparation). 
Alternatively, the upper limit of winding angle $\sim 1$ also yields a comparable resistance $\Lambda_1\approx 0.25 v/(10^5 \,{\rm m/s}) \,{\rm ohm}$ \citep{lai2012}. When the star-planet system approaches a synchronous state, i.e., $v\rightarrow 0$, $\Lambda_1$ tends to vanish and hence the circuit will be established through the planet's atmosphere instead of through Alfv\'en \citep{laine2012}. The resistance of planet's atmosphere arising from Na$^+$ and K$^+$ is about $\Lambda_2\approx 0.1$ ohm \citep{laine2008, french2012}. We can write $\Lambda=\Lambda_1+\Lambda_2$ to take into account both the synchronous and asynchronous states.

At an asynchronous state, inserting $\Lambda\approx\mu v/C_d$ into (Eq. \ref{unipolar-torque}) we obtain the torque magnitude $\Gamma\approx C_d \cdot a \cdot 4R_{\rm obs}^2 \cdot P_m$, where $P_m=\left[B_\star(R_\star/a)^3\right]^2/\mu$ is employed. It is interesting that the magnetic torque of unipolar induction is almost identical to that of dipole-dipole interaction $\Gamma\approx C_d \cdot a \cdot \pi R_{\rm obs}^2 \cdot P_t$ under the circumstance $P_t\approx P_m$. The reason is that the dipole-dipole interaction is caused by magnetic reconnection and the Alfv\'en wing in unipolar interaction is also induced by the same physical mechanism. This is also shown in \citet{zarka2007} that the two magnetic dissipations are comparable to each other.

\subsubsection{Planet-moon and star-moon unipolar interactions}
In the planet-moon interaction, when the moon is unmagnetized, e.g., 
Jupiter-Io, $R_{\rm obs}$ for the star-planet unipolar induction 
\eqref{unipolar-torque} should be replaced with the physical size 
of exomoon, i.e., moon's radius $R_m$. The resistance 
$\Lambda_{p, m}=1/(\sigma_{p, m} R_{p, m})$ arises from a series 
circuit through the moon itself and across the footprint of the 
planetary field. Magnitude of moon's intrinsic conductivity $\sigma_m$ 
depends on its composition and structure, which varies over many 
order of magnitude between volatile ices, condensed or molten silicates, 
and iron.  Many known super Earths have density comparable to that of 
the Earth and surface temperature above the melting point of silicates. 
If the surface of exomoons around hot Jupiters (with orbits similar to 
these close-in super Earths) is heated to the same temperature and is 
covered with magma ocean, their conductivity may be very large. But around 
warm and cold Jupiters, icy moons' intrinsic conductivity may be much smaller.

In highly conductive moons with negligible magnetic diffusivity, the 
field distortion induced by the planet-moon differential rotation is 
quickly amplified. Eventually, the induced field diffuses through the 
surrounding plasma in the Alfv\'en-wing wake of the moon's orbit (\S\ref{sec:starplanetunip}).  
Around Jupiter, Io's effective conductivity is caused by the plasma 
torus around its orbit\citep{Neubauer1980}. Hot Jupiters and their 
moons are also embedded in the plasma from their host stars' wind
\citep{zarka2007, strugarek2016}.  
In view of these possibilities and uncertainties, we adopt Io's 
conductivity $\sigma_m\approx 10^{-8}\mbox{ ohm}^{-1} \mbox{ cm}^{-1}$ and the typical radius 
$R_m\approx 3\times 10^8\mbox{ cm}$ so that moon's effective 
resistance is $1/(\sigma_mR_m)\approx 1$ ohm, and the resistance 
at the foot of the planetary field is also around 1 ohm 
\citep{goldreich1969}. Our adopted fiducial value may be
an overestimate for icy moons' effective $\sigma_m$ and therefore enhance
the probability of their retention around warm and cold Jupiters.
We set $\Lambda=1/(\sigma_mR_m)$ in \eqref{unipolar-torque}, and consequently the 
magnetic torque due to the planet-moon unipolar induction reads
\begin{equation}\label{moon-magnetic-torque}
\Gamma^m_{pm} \approx 4\sigma_m(\omega_m-\Omega_p)a_m^2R_m^3B_p^2(R_p/a_m)^6
\end{equation}
where $\omega_m$ is moon's orbital frequency and $a_m$ its orbital semi-major axis. 
The subscript $p$ denotes planet (primary) and $m$ moon (secondary), and the 
superscript $m$ magnetic torque.

In addition to the planet-moon unipolar interaction, star and moon can also induce unipolar interaction. In star's frame of reference, the electric field induced by star-moon unipolar interaction is $\bm E=(\bm v_p+\bm v_m)\times\bm B)$ where $\bm v_p$ is planet's velocity relative to star, $\bm v_m$ is moon's velocity relative to planet, and $\bm B$ is star's magnetic field at planet (precisely speaking at moon but the upper limit of planet-moon distance is 10\% of star-planet distance, see \S\ref{sec:radii}). The Lorentz force is then $\bm F\propto -(\bm v_p+\bm v_m)B^2$. The force arm is $\bm r_p+\bm r_m$ where $\bm r_p$ is the position vector of planet relative to star and $\bm r_m$ the position vector of moon relative to planet. We now calculate the torque $\bm\Gamma^m _{\star m}=(\bm r_p+\bm r_m)\times\bm F\propto \bm r_p\times\bm v_p+\bm r_p\times\bm v_m+\bm r_m\times\bm v_p+\bm r_m\times\bm v_m$. What we are concerned with is the moon's orbital evolution around planet, and therefore, the first term $\bm r_p\times\bm v_p$ arising from the star-planet interaction should be subtracted, the second and third terms proportional to $\cos(\omega_m-\omega_p)t$ vanish when averaged over the moon's orbit around planet, and eventually only the last term $\bm r_m\times\bm v_m$ contributes to the moon's orbital evolution. We can readily write the torque of star-moon unipolar interaction
\begin{equation}\label{eq:torquestarmoon}
\Gamma^{m}_{\star m}\approx 4\sigma_m(\omega_m-\Omega_p)a_m^2R_m^3B_\star^2(R_\star/a_p)^6.
\end{equation}
Compared to the torque of planet-moon unipolar interaction \eqref{moon-magnetic-torque}, the difference lies in magnetic field, i.e., planetary field at moon in the former whereas stellar field at planet in the latter. Both $\Gamma^m_{p m}$ \eqref{moon-magnetic-torque} and $\Gamma^{m}_{\star m}$ \eqref{eq:torquestarmoon} are in the direction of the moon's motion relative to the planet. Our estimation shows that for a hot-Jupiter system the star-moon torque is stronger than the planet-moon torque whereas for a cold-Jupiter system the former decaying as $a_p^{-6}$ is far weaker than the latter.

\subsection{Combination of tidal and magnetic torques}\label{sec:all-torques}

\begin{figure}
\centering
\includegraphics[scale=0.18]{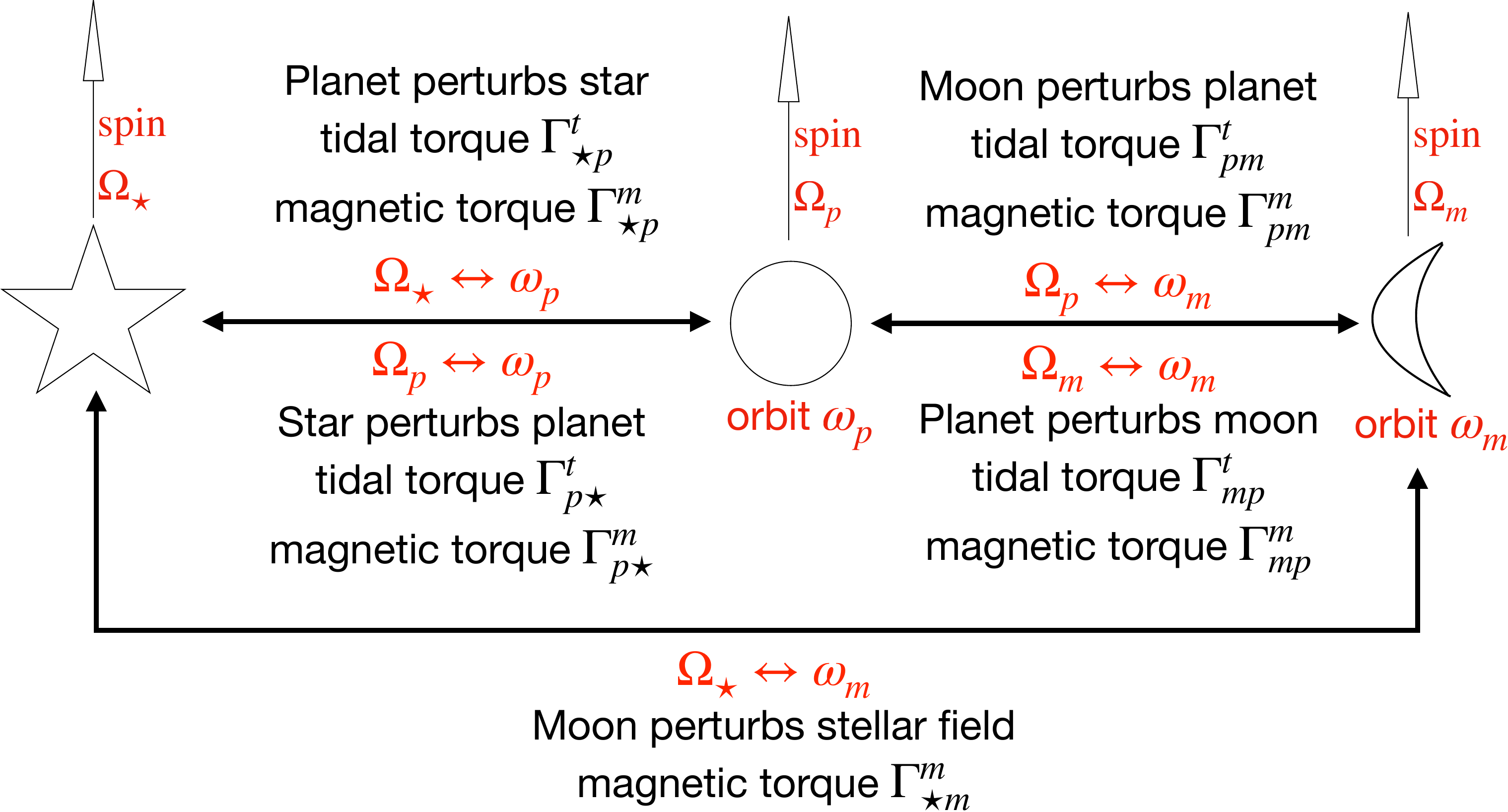}
\caption{Schematic illustration of tidal and magnetic torques in a star-planet-moon system. The first subscript denotes the primary, the second subscript the secondary, and the superscript the type of torque (tidal or magnetic).}\label{sketch}
\end{figure}

In a two-body system, e.g., star-planet or planet-moon system, both tidal and magnetic torques can transfer angular momentum between the orbit and the spins. We will summarize all the torques in this subsection. Figure \ref{sketch} shows the sketch of tidal and magnetic torques in a star-planet-moon system. The arrows depicts the primary and secondary, i.e., arrows pointing from secondary to primary. Here the symbols are defined as follows: the first subscript denotes the primary on which the torque is exerted, the second subscript denotes the secondary which exerts the torque (`$\star$' for star, `$p$' for planet and `$m$' for moon), and the superscript denotes the type of torque (`$t$' for tidal torque and `$m$' for magnetic torque).

\subsubsection{The Q-value and tidal time lag for equilibrium and dynamical tides}
The tidal torque due to the large-scale equilibrium tide in convection zone is proportional to the tidal frequency $\omega$, i.e., the orbital frequency relative to the primary's spin frequency, and also proportional to turbulent viscosity $\nu_t$ in primary's convection zone. The combined effect of tidal frequency and turbulent viscosity on tidal torque is modelled with the tidal quality factor $Q=E_0/\oint\dot Edt\simeq E_0/\dot EP\simeq E_0/\Gamma^t$ where $\oint$ denotes the integral over an orbital cycle, $E_0$ is the tidal deformation energy, namely the product of tidal force $\simeq GM_1M_2R_1/a^3$ and tidal deformation $\simeq (M_2/M_1)(R_1^4/a^3$), $\dot E$ is tidal dissipation rate, $P\simeq 1/\omega$ is tidal period, and $\Gamma^t$ is tidal torque. Thus, the tidal torque is $\Gamma^t\simeq GM_2^2R^5/(a^6Q)$.

Usually $Q$ can be written as $Q\simeq 1/\omega\tau$ where $\tau=\nu_t R_1/GM_1$ is the tidal time lag \citep{hut1981, eggleton1998}. In the weak friction theory, turbulent viscosity $\nu_t$ or tidal time lag $\tau$ is independent of tidal frequency $\omega$ such that tidal torque $\Gamma^t$ is proportional to tidal frequency $\omega$; in Zahn's theory \citep{zahn1977}, turbulent viscosity is inversely proportional to tidal frequency such that tidal torque is independent of tidal frequency; in Goldreich's theory \citep{goldreich1977} turbulent viscosity is inversely proportional to the square of tidal frequency such that tidal torque is inversely proportional to tidal frequency. The most recent numerical simulations support Goldreich's theory \citep{barker2020a, barker2020b}.

In addition to equilibrium tide, dynamical tide (e.g., inertial waves in convection zone or internal gravity waves in radiation zone) also induces tidal dissipation and hence tidal torque. The tidal torque induced by dynamical tide, especially by inertial waves in convection zone, dominates over that induced by equilibrium tide \citep{ogilvie-lin-2004, ogilvie2014}. The tidal torque induced by inertial wave is strongly frequency-dependent because inertial wave has a very dense spectrum \citep{ogilvie-lin-2004, ogilvie2014}. To avoid this complexity, tidal torque induced by dynamical tide is usually calculated with its frequency-average and also modelled with $Q$ number \citep{ogilvie2014, strugarek2017}. However, tidal torque will sharply decay to zero when tidal frequency approaches zero at a synchronous state \citep{ogilvie-lin-2004, ogilvie-lin-2007}. In order to take into account the synchronous state we calculate the tidal torque induced by dynamical tide with tidal time lag $\tau$ instead of tidal $Q$,
\begin{equation}
\Gamma^t\simeq GM_2^2R^5\omega\tau/a^6.
\label{eq:tidallag}
\end{equation}
The superscript $t$ denotes tidal torque. The observations of close-in binary circularization \citep{mathieu1994}, exoplanet's \citep{penev2018} and Galilean moons' migration \citep{yoder1981, lainey2009} suggest that star's or planet's tidal $Q$ is about $Q\sim 10^{5-6}$. The tidal frequency is about $\omega\simeq 10^{-5}~{\rm s}^{-1}$, and hence tidal time lag of dynamical tide can be estimated $\tau\sim 1-10~{\rm s}$.

\subsubsection{Bypassing the valleys of dynamical tides}
As discussed in the last subsection \S\ref{sec:magnetic-interaction}, the magnetic torque, in addition to tidal torque, can cause angular momentum transfer between the host star and its planet or between the host planet and its moon. For a more complete view, 
we introduce a fiducial model for a young (a few Myr old) planetary system with the typical parameters listed in Table \ref{table2} 
(the details of calculation will be shown in the next section).   
For this model, we evaluate all the eight torques in a star-planet-moon system (first row in
Table \ref{table3}) as well the corresponding spin-orbit interchange due to the angular 
momentum transfer by each torque (second row). The third row shows 
the values of the torques at an asynchronous state.  We consider only a circular orbit such that there is no interchange 
of spin and eccentricity. If the moon is un-magnetized, the last torque $\Gamma_{mp}^m$ induced by planet on moon's magnetic 
field would be irrelevant.

\begin{table*}
\centering
\begin{tabular*}{0.9\textwidth}{@{\extracolsep{\fill}}|c|c|c|c|c|c|c|c|c|c|c|c|c|c|c|}
\hline
$M_\star$ & $R_\star$ & $P_\star$ & $B_\star$ & $\tau_\star$ & $M_p$ & $R_p$ & $P_p$ & $B_p$ & $\tau_p$ & $M_m$ & $R_m$ & $\rho_m$ & $a_p$ & $a_m$ \\ \hline
$M_\odot$ & $2R_\odot$ & 3d & 1500G & 1s & $M_J$ & $2R_J$ & 40h & 100G & 1s & $10^{26}$g & $3\times 10^8{\rm cm}$ & $3~{\rm g/cm^3}$ & 0.05au & $6R_J$ \\ \hline
\end{tabular*}
\caption{The typical parameters of a young star-planet-moon system. In this fiducial model, the plasma number density $N=5(t/4.6{\rm Gyr})^{-2}(a_p/1{\rm au})^{-2}{\rm cm^{-3}}$. The coefficient of planet's moment of inertia $\alpha_p=0.1$ and moon's conductivity $\sigma_m=10^{-8}{\rm ohm^{-1}cm^{-1}}$.}\label{table2}
\end{table*}

\begin{table*}
\centering
\begin{tabular*}{0.9\textwidth}{@{\extracolsep{\fill}}|c|c|c|c|c|c|c|c|c|}
\hline 
$\Gamma_{\star p}^t$ & $\Gamma_{\star p}^m$ & $\Gamma_{p\star}^t$ & $\Gamma_{p\star}^m$ & $\Gamma_{pm}^t$ & $\Gamma_{pm}^m$ & $\Gamma_{mp}^t$ & $\Gamma_{mp}^m$ & $\Gamma^{m}_{\star m}$ \\ \hline
$\Omega_\star\leftrightarrow\omega_p$ & $\Omega_\star\leftrightarrow\omega_p$ & $\Omega_p\leftrightarrow\omega_p$ & $\Omega_p\leftrightarrow\omega_p$ & $\Omega_p\leftrightarrow\omega_m$ & $\Omega_p\leftrightarrow\omega_m$ & $\Omega_m\leftrightarrow\omega_m$ & $\Omega_m\leftrightarrow\omega_m$ & $\Omega_\star\leftrightarrow\omega_m$ \\ \hline
$5.2\times10^{32}$ & $1.7\times10^{33}$ & $2.1\times10^{34}$ & $2.4\times10^{31}$ & $7.8\times10^{25}$ & $3.1\times10^{25}$ & $1.8\times10^{30}$ & N/A & $2.2\times10^{26}$ \\ \hline
\end{tabular*}
\caption{The first row shows all the torques in the star-planet-moon system. The second row shows the corresponding spin-orbit interchange. The third row shows their values (erg) at an asynchronous state with the typical parameters listed in Table \ref{table2}.}\label{table3}
\end{table*}

Furthermore, as discussed in \S\ref{sec:magnetic-interaction}, magnetic torque is proportional to tidal frequency. That is, contrary to the torque due to dynamical tides associated with inertial waves, magnetic torque varies smoothly with tidal frequency. Therefore, \textit{\textbf{during the orbital evolution when tidal frequency varies, magnetic torque smooths the total torque in the sense that it compensates tidal torques and bypasses frequency valleys where dynamical-tide response is ineffective at those particular tidal frequencies}}. 

\begin{figure}
\centering
\includegraphics[scale=0.5]{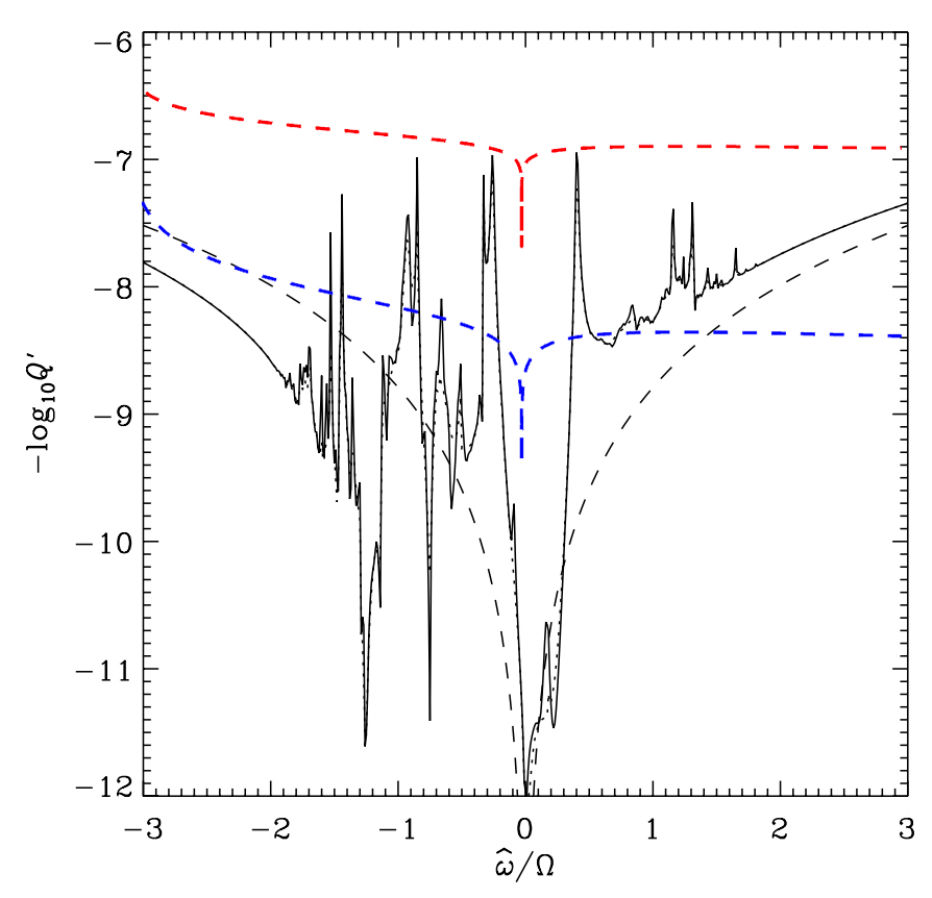}
\caption{Illustration of the smoothing effect of magnetic torque on dynamical tide. Black curves are adapted from \citet{ogilvie-lin-2007}: stellar tidal dissipation versus tidal frequency, the vertical axis $Q'\approx Q/k$ denotes tidal dissipation (where the apsidal motion constant $k\approx 0.035$ for the Sun), and the horizontal axis denotes the tidal frequency (i.e., the difference between orbital and stellar spin frequencies). The dashed black lines denote the results of equilibrium tide. The two dashed color lines show the magnetic torques of the system of a solar-like star with a hot Jupiter: red denotes stellar surface field 1000 gauss and planetary surface field 100 gauss (young planetary system) and blue denotes stellar surface field 10 gauss and planetary surface field 10 gauss (mature planetary system). Magnetic torque compensates the minuscule tidal torque at some frequencies, and higher fields correspond to a stronger magnetic torque.}\label{fig:magnetic-torque}
\end{figure}

Figure \ref{fig:magnetic-torque} shows the illustration of this smoothing effect of magnetic torque on dynamical tide. In the Figure the equivalent tidal $Q$ number due to magnetic torque is estimated with $Q\simeq E_0/\Gamma^m$ where $E_0$ is tidal deformation energy and $\Gamma^m$ is magnetic torque. It should be noted that the equivalent magnetic $Q\propto a^{-3}$($E_0\propto a^{-6}$ and $\Gamma^m\propto a^{-3}$), such that magnetic $Q$ in Figure \ref{fig:magnetic-torque} is asymmetric about the tidal frequency $\hat\omega=0$ (in Figure adapted from \citet{ogilvie-lin-2007} tidal frequency is denoted by Doppler-shifted $\hat\omega$ rather than $\omega$). We can then infer such a situation: when a planet migrates inward, say, in a disk due to the angular momentum transfer between the planet and the disk, tidal torque becomes very weak at some orbital frequencies. However, magnetic torque continues to induce the inward migration until the planet plunges into the star.

\subsubsection{A brief synopsis of tidal and magnetic torques}
To wrap up this subsection, we summarize the expressions of all the torques already discussed:
\begin{equation}
\Gamma_{\star p}^t \approx GM_p^2R_\star^5(\omega_p-\Omega_\star)\tau_\star/a_p^6, \nonumber
\end{equation}
\begin{equation}
\Gamma_{p\star}^t \approx GM_\star^2R_p^5(\omega_p-\Omega_p)\tau_p/a_p^6, \nonumber
\end{equation}
\begin{equation}
\Gamma_{\star p}^m \approx 4(\omega_p-\Omega_\star)(B_\star^2/\Lambda)a_p^2R_p^2(B_p/B_\star)^{2/3}(R_\star/a_p)^4, \nonumber
\end{equation}
\begin{equation}
\Gamma_{p\star}^m \approx 4(\omega_p-\Omega_p)(B_p^2/\Lambda)a_p^2R_\star^2(B_\star/B_p)^{2/3}(R_p/a_p)^4, \nonumber
\end{equation}
\begin{equation}
\Gamma_{pm}^t \approx GM_m^2R_p^5(\omega_m-\Omega_p)\tau_p/a_m^6, \nonumber
\end{equation}
\begin{equation}
\Gamma_{mp}^t \approx GM_p^2R_m^5(\omega_m-\Omega_m)\tau_m/a_m^6, \nonumber
\end{equation}
\begin{equation}
\Gamma_{pm}^m \approx 4\sigma_m(\omega_m-\Omega_p)a_m^2R_m^3B_p^2(R_p/a_m)^6, \nonumber
\end{equation} 
\begin{equation}
\Gamma^{m}_{\star m} \approx 4\sigma_m(\omega_m-\Omega_p)a_m^2R_m^3B_\star^2(R_\star/a_p)^6. \nonumber
\end{equation}

The positive sign means that orbital motion is faster than spin so that the torque spins up the primary body denoted by the first subscript; and vice versa, the negative sign corresponds to the spindown of the primary body. The typical values of the torques are calculated with the parameters in Table \ref{table2} and listed in the third row of Table \ref{table3}. The tidal torque $\Gamma_{mp}^t$ raised by planet on moon depends on moon's tidal quality factor $Q_m=1/(\omega_m-\Omega_m)\tau_m$, assumed to be 67, i.e., Io's $Q$ \citep{lainey2009}. $\Gamma_{mp}^t$ is estimated to be stronger than the tidal torque $\Gamma_{pm}^t$ or magnetic torque $\Gamma_{pm}^m$ raised by moon on planet. However, since moon's mass is very small, its spin and orbit will quickly reach a synchronous state under such a strong torque, therefore, $\Gamma_{mp}^t$ will soon vanish, and moon's orbital migration will then be controlled by $\Gamma_{pm}^t$ and $\Gamma_{pm}^m$. 

\subsection{Relative strength of magnetic versus tidal torques for planets and moons}
\label{sec:magtide}

In an \textit{\textbf{asynchronous}} state, $\Gamma_{p\star}^t\approx GM_\star^2R_p^5/(a_p^6Q_p)$ and $\Gamma_{\star p}^t\approx GM_p^2R_\star^5/(a_p^6Q_\star)$. Suppose that the two tidal $Q$'s are comparable, and hence the ratio is $\Gamma_{p\star}^t/\Gamma_{\star p}^t\approx (M_\star/M_p)^2(R_p/R_\star)^5\approx 10$. Consequently, $\Gamma_{p\star}^t$ leads to the synchronization of the planet's spin before it undergoes significant migration.  As $\Omega_p$ approaches $\omega_p$, 
$\Gamma_{\star p}^t$ dominates over the diminishing but finite $\Gamma_{p\star}^t$ and the planet migrates with an evolving 
$\omega_p$ and $\Omega_\star$ while a state of near $\Omega_p - \omega_p$ synchronism is being maintained.
In addition, $\Gamma_{\star p}^m$ (which is far stronger than $\Gamma_{p\star}^m$) also contributes to the planet's migration.
Similarly, $\Gamma_{pm}^t$ and $\Gamma_{pm}^m$ leads to the moon's near spin-orbit ($\Omega_m-\omega_m$) synchronism.  The leading torques for moon's 
migration are tidal torque $\Gamma_{pm}^t$ and magnetic torques $\Gamma_{pm}^m$ ($\Gamma^{m}_{\star m}$ is important for the 
close star-planet distance but decays quickly as $a_p^{-6}$). 

\begin{figure*}
\centering
\includegraphics[scale=0.5]{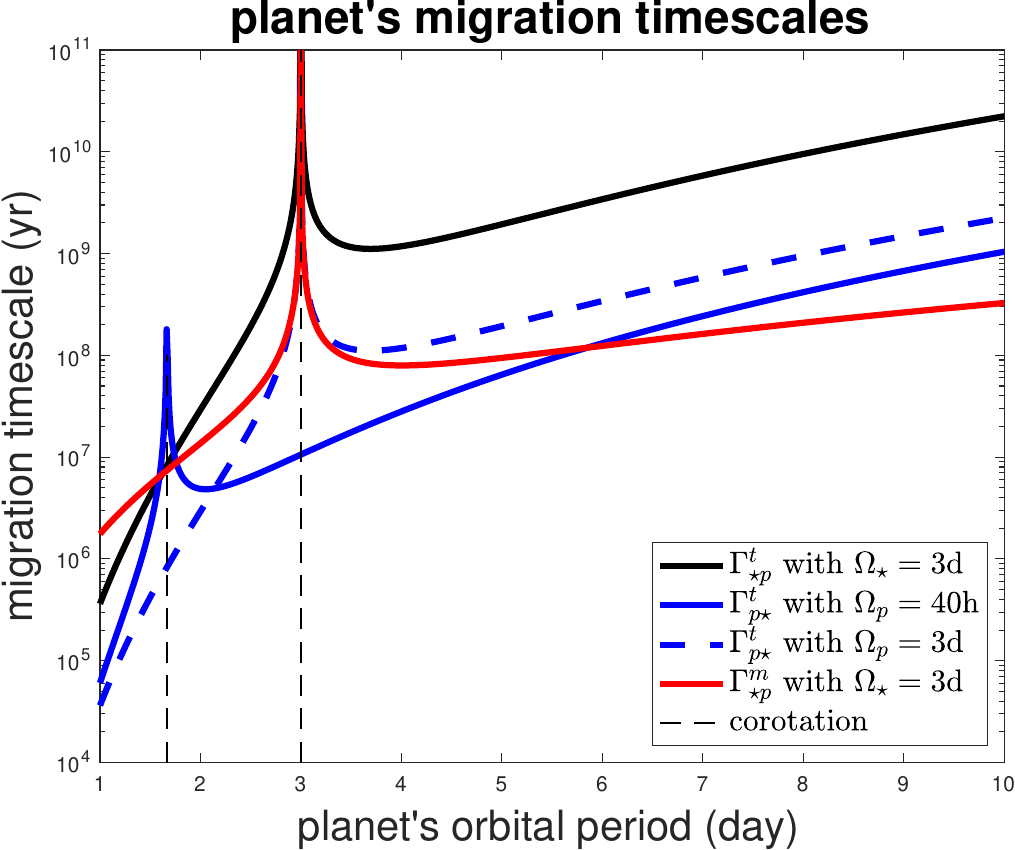}
\includegraphics[scale=0.5]{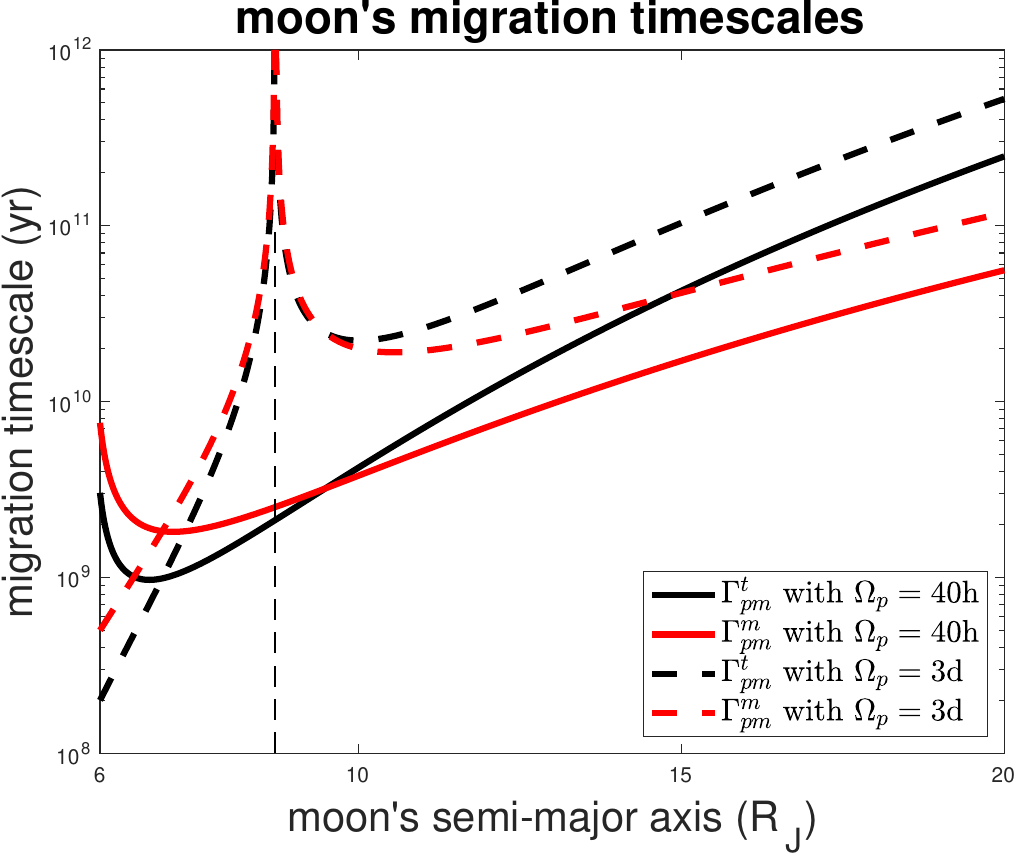}
\caption{Left panel: planet's migration timescales versus planet's orbital period due to different torques and planet's spin periods (40 hours and 3 days). Star's spin period is 3 days. Right panel: moon's migration timescales versus moon's semi-major axis due to different torques and planet's spin periods.
}\label{migration}
\end{figure*}

Based on the model parameters for a fiducial young system (Table \ref{table2}), we now compare the relative strength of the torques for planet's or moon's migration. Figure \ref{migration} shows the tidal and magnetic migration timescales for planet's migration (left panel) and for moon's migration (right panel). 
We firstly focus on planet's migration (left panel). The comparison between the two tidal torques $\Gamma_{\star p}^t$ and $\Gamma_{p\star}^t$ shows that in an asynchronous state $\Gamma_{p\star}^t$ is stronger than $\Gamma_{\star p}^t$ by one order, which is consistent with our analysis. The comparison between the tidal torque $\Gamma_{\star p}^t$ and magnetic torque $\Gamma_{\star p}^m$ shows that the magnetic torque always wins out the tidal torque both inside and outside corotation radius. The magnetic torque $\Gamma_{\star p}^m$ wins out $\Gamma_{p\star}^t$ when planet's orbital period becomes longer than 6 days, because the former decays as $a_p^{-2}$ whereas the latter $a_p^{-6}$. However, $\Gamma_{\star p}^m$ decreases as the stellar field decays and radius contracts on a timescale of a few $10^{7}$ yr.   
Around mature main sequence stars (with a life span of a few Gyrs), neither the magnetic nor tidal torque can significantly 
influence the orbits of planets with period $\gtrsim 10$ days.

Next we move to moon's migration (right panel). Outside the corotation radius, the magnetic torque $\Gamma_{pm}^m$ wins out the tidal torque $\Gamma_{pm}^t$ if the moon is located outside 10 $R_J$.
However, the migration timescale is already longer than 1 Gyr.  Over such a long timescale, $\Gamma_{pm}^m$ would be much reduced as $B_p$ declines and $R_p$ contracts.  Thus, magnetic field is unlikely to have a dominant and significant influence on moon's migration.

\section{Retention of exomoons}\label{sec:moon}

Recently, exomoons are possibly found in the circumplanetary disk of, for example, PDS70c \citep{benisty2021}, and the retention of exomoons has now been studied, e.g., \citep{Tokadjian2020}. In this section, we will study the circumplanetary disk and the timescales of planet-moon dynamics, and then investigate under what situations the exomoon can retain.

\subsection{Inner and outer radii of Circumplanetary disks}\label{sec:radii}

\begin{figure}
\centering
\includegraphics[scale=0.6]{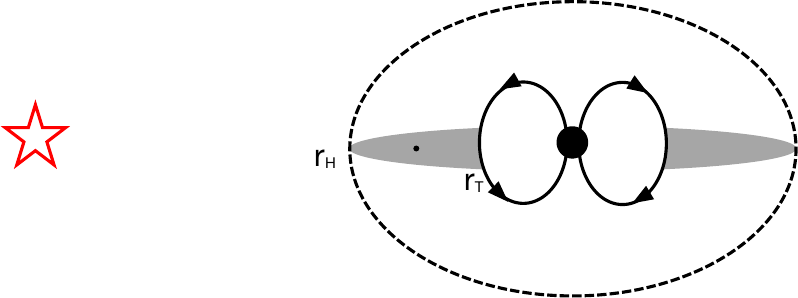}
\caption{Sketch of inner and outer radii of circumplanetary disk (grey area).}\label{sketch1}
\end{figure}

\begin{figure*}
\centering
\includegraphics[scale=0.5]{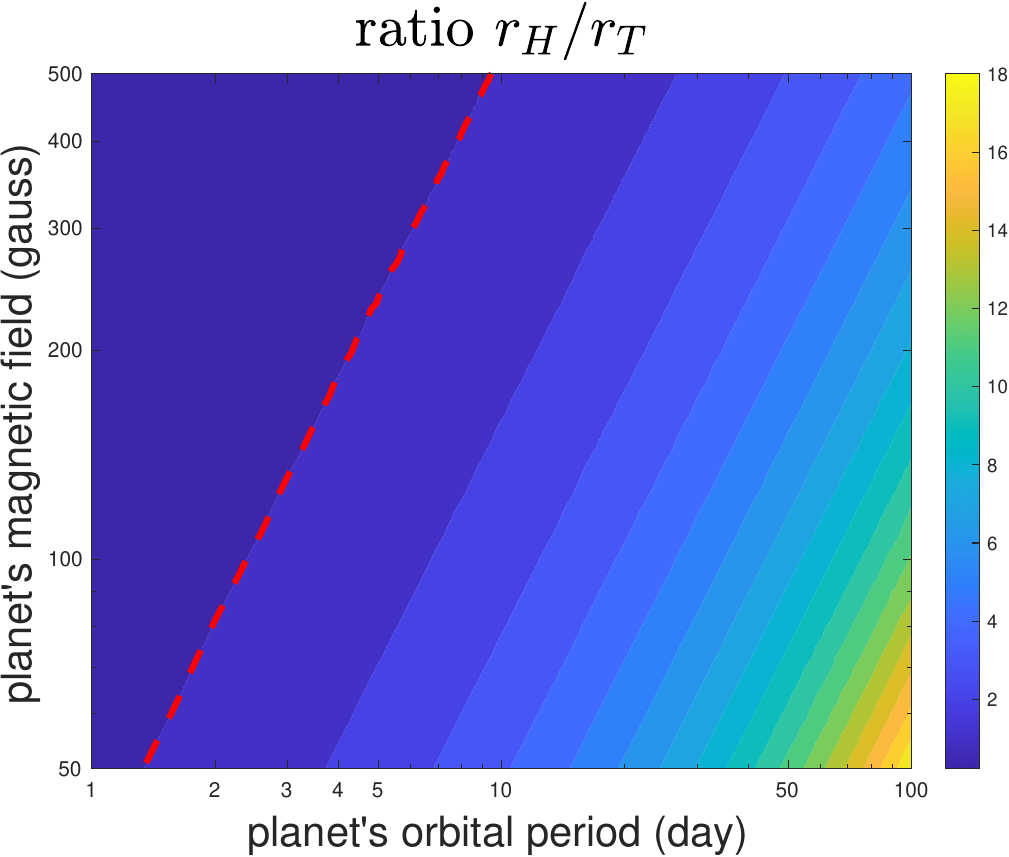}
\caption{Contours for the ratio $r_H/r_T$ of the two radii of circumplanetary disk. The red dashed line denotes the ratio $r_H/r_T=1$. The area to the right of the red dashed line with $r_H/r_T>1$ is favourable for moon's survival and the area to the left with $r_H/r_T<1$ is unfavourable for moon's survival.}\label{rTrH}
\end{figure*}

The radius of circumplanetary disk cannot exceed the planetary Hill radius, otherwise gas would be caught by host star \citep{martin2011} 
\begin{equation}
r_H\approx(M_p/M_\star)^{1/3}a_p\approx 0.1 a_p,
\end{equation} 
where $a_p$ is the planetary orbital semi-major axis. Hill radius $r_H$ is considered to be the outer radius of circumplanetary disk. At the innermost circumplanetary disk, planet's magnetic field is so strong that it can truncate the disk. The magnetic truncation radius, which is considered to be the inner radius of circumplanetary disk, can be estimated by the balance between magnetic pressure and kinetic pressure $\left[B_p(R_p/r_T)^3\right]^2/(2\mu)\approx\rho v^2$ (or equivalently the balance between Alfv\'en speed and orbital speed) where density $\rho$ is estimated with the accretion rate of circumplanetary disk $\rho=\dot M_p/(4\pi r_T^2v)$ and velocity $v$ follows Keplerian motion $v=(GM_p/r_T)^{1/2}$. The magnetic truncation radius can be estimated 
\begin{equation}
r_T\approx(B_p^4R_p^{12}/GM_p\dot M_p^2)^{1/7}
\label{eq:rtrun}
\end{equation}
\citep{koenigl1991}.  Inserting the typical values $M_p=M_J$, $R_p=2R_J$, $\dot M_p=10^{-7}M_J/yr$ and $B_p=100$ gauss, we obtain $r_T\approx 3R_p\approx 6R_J$, which is exactly the Io's position. If spin-orbit synchronism can be established at $r_T$, Jupiter would spin with a period 3-4 times longer than its present-day value, and the subsequent (a factor of 2) contraction would increase its spin rate to its present-day value. The magnetic truncation radius $r_T$ gives the lower bound of the inner radius of circumplanetary disk \citep{koenigl1991, batygin2018}, whereas the Hill radius gives the upper bound of the outer radius \citep{lin1976, papaloizou1977, machida2008, fung2015, li2021}. 
Compared to $r_H\approx 0.1a_p$, $r_T\approx 6R_J$ is not much less than $r_H$ for a hot Jupiter with $a_p$, say, 0.05 au. Since $r_T\propto B_p^{4/7}$, if the young planetary field reaches, say, 300 gauss, then $r_T>r_H$ such that the circumplanetary disk does not exist and hence the exomoon cannot be born. Figure \ref{sketch1} shows the schematic configuration of star-planet-moon system, where moon lies in circumplanetary disk between inner radius $r_T$ and outer radius $r_H$. Figure \ref{rTrH} shows the contours of ratio of the two radii $r_H/r_T$ versus planet's orbital period and magnetic field. The area to the right of the red dashed line with the ratio greater than 1 is favorable for moon's retention. We draw a conclusion that \textbf{\textit{a planet with too strong magnetic fields or too short distance from its host star tends not to have exomoons}}.

\subsection{Evolution of planet's corotation radius and moon's migration}\label{sec:migration}

\begin{figure}
\centering
\includegraphics[scale=0.5]{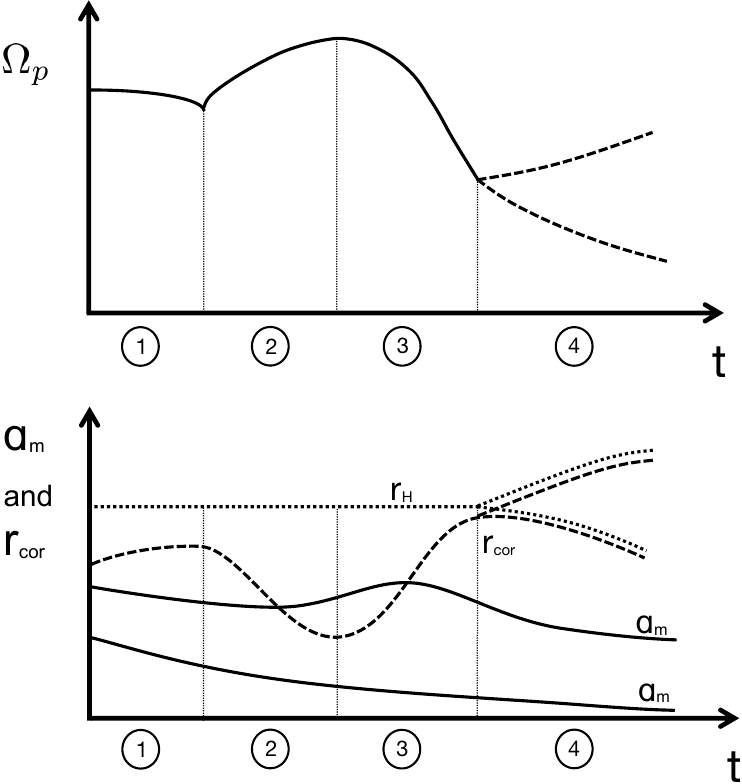}
\caption{Schematic illustration of a close-in planet's spin evolution (top), moon's semi-major axis $a_{\rm m}$ and planet's corotation 
radius $r_{\rm cor}$ evolution (bottom). Stage \textcircled{1}: planet slightly inflates; stage \textcircled{2}: planet contracts, 
stage \textcircled{3}: planet spins down due to tidal interaction with its host star, stage \textcircled{4}: planet migrates inward (spinup) 
or outward (spin-down).  
%The evolution of $r_{\rm cor}$ and $a_m$ would terminate at the end of stage \textcircled{2}.
}\label{sketch2}
\end{figure}

When a gas giant planet is born, it spins with an angular frequency comparable to the Keplerian frequency at $r_T$. On the Kelvin-Helmholtz thermal timescale of Gyr, it spins up due to its contraction. However, at its young age a close-in planet spins down due to the tidal torque $\Gamma_{p\star}^t$ raised by its host star. 
During the planet's spin-down, the corotation radius of any circumplanetary satellites moves outward. The top panel 
of Figure \ref{sketch2} shows planet's spin evolution. In stage \textcircled{1} shortly after planet is born, planet 
inflates slightly; in stage \textcircled{2} planet contracts such that it spins up; in stage \textcircled{3} planet 
spins down due to star-planet tidal interaction; and in stage \textcircled{4} planet inward migrates and spins up or 
outward and down. The stage \textcircled{3} when a close-in planet spins down due to star-planet interaction and 
corotation radius moves out is what we are concerned with.  But the spin of a long-period (with $\tau_{\rm cor} > 
\tau_\star$) the planet does not spin down significantly throughout its host star's life span and its evolution terminates
at the end of stage \textcircled{2}.

In the meanwhile, a moon located outside the corotation radius also migrates outward due to the tidal and magnetic torques $\Gamma_{pm}^t$, $\Gamma_{pm}^m$ and $\Gamma^{m}_{\star m}$. If the corotation radius moves faster than moon's outward migration, the moon will be located inside the corotation radius, migrate inward and eventually plunge into the planet to form a planetary ring \citep{toi3757b2022}. The bottom panel of Figure \ref{sketch2} shows moon's migration and planet's corotation radius evolution. Solid curves represents the three possible moon migration tracks, and dashed curve planet's corotation radius evoltuion corresponding to its spin evolution in the top panel. 

In the first place we estimate the planet's spin frequency $\Omega_p$ when it forms. We assume that the planet's corotation radius $r_{\rm cor}$ is located at its magnetic truncation radius $r_T$ \citep{batygin2018, ginzburg2020, hasegawa2021}. Using the expression of $r_T$ in the last subsection, we readily derive the planet's spin frequency $\Omega_p=(GM_p/r_T^3)^{1/2}\approx 0.1(GM_p/R_p^3)^{1/2}$, which is $10\%$ of the break-up frequency and consistent with the observations \citep{bryan2020}. With the typical values, the planet's spin period is about 40 hours.

In addition to the spin-up on the Kelvin-Helmholtz timescale due to contraction, a hot Jupiter can spin down on a much shorter timescale induced by the tidal torque $\Gamma_{p\star}^t$ exerted on the planet's host star. With the planet spinning down, the corotation radius of satellites moves outward on a timescale 
\begin{equation}\label{tau1}
\tau_{\rm cor}=r_{\rm cor} / {\dot r}_{\rm cor} \approx 1.5\alpha_pM_pR_p^2\Omega_p/|\Gamma_{\rm p \star}^t|
\end{equation}
where $\alpha_p\approx 0.1$ is the coefficient of planet's moment of inertia and the coefficient 1.5 arises from Kepler's third 
law $\Omega_p=\omega_m\propto a_m^{-1.5}$. 
\begin{comment}
{\color {red} Not sure if that factor of 1.5 is justified since the star's tidal torque is applied
on the planet's spin $\Omega_{\rm p}$ not on the moon's orbit ($\omega_{\rm m}$).}  {\color {red} Although at $\Gamma_{\rm p \star} ^t$
vanishes when the planet's spin approaches corotation with its orbit around its host star (i.e. $\Omega_p=\omega_p$),
$\omega_p \propto a_p^{-3/2}$ continues to evolve with the planet's orbit due to its tidal torque on its host star.  
Even after the planet has achieved a state of total synchronism with its host star ($\Omega_p=\omega_p=\Omega_\star$), 
the system continues to evolve due to stellar-wind induced angular momentum loss.  We need to be careful about its application to Fig. 7.}
\end{comment}

An exomoon outside/inside its host planet's corotation radius migrates outward/inward due to the tidal torque $\Gamma_{pm}^t$ and the magnetic torque $\Gamma_{pm}^m$ (these two torques are comparable) as well as $\Gamma^{m}_{\star m}$ (this torque is important for a hot-Jupiter system). The moon's migration timescale is estimated as
\begin{equation}
\tau_{\rm mig}= a_m/{\dot a}_m \approx \frac{0.5 M_m(GM_pa_m)^{1/2}}{|\Gamma_{pm}^t+\Gamma_{pm}^m+\Gamma^{m}_{\star m}|}.
\label{eq:taumig}
\end{equation}

At the end of stage \textcircled{2}, the moon would be retained with an expanding orbit if its $a_m > r_{\rm cor}$.
In the limit $a_m < r_{\rm cor}$, the moon's would decay into the planet if $\tau_{\rm cor} > \tau_{\rm mig}$, but 
it could be retained, at least until the end of stage \textcircled{3} if $\tau_{\rm cor} < \tau_{\rm mig}$.  
Thereafter, the moon would still plunge into the planet if $a_m < r_{\rm cor}$ and $\tau_{\rm mig} \leq \tau_\star$ before the
host star evolves off the main sequence on a timescale $\tau_\star$.  Otherwise it would be retained.

\begin{figure*}
\centering
\includegraphics[scale=0.5]{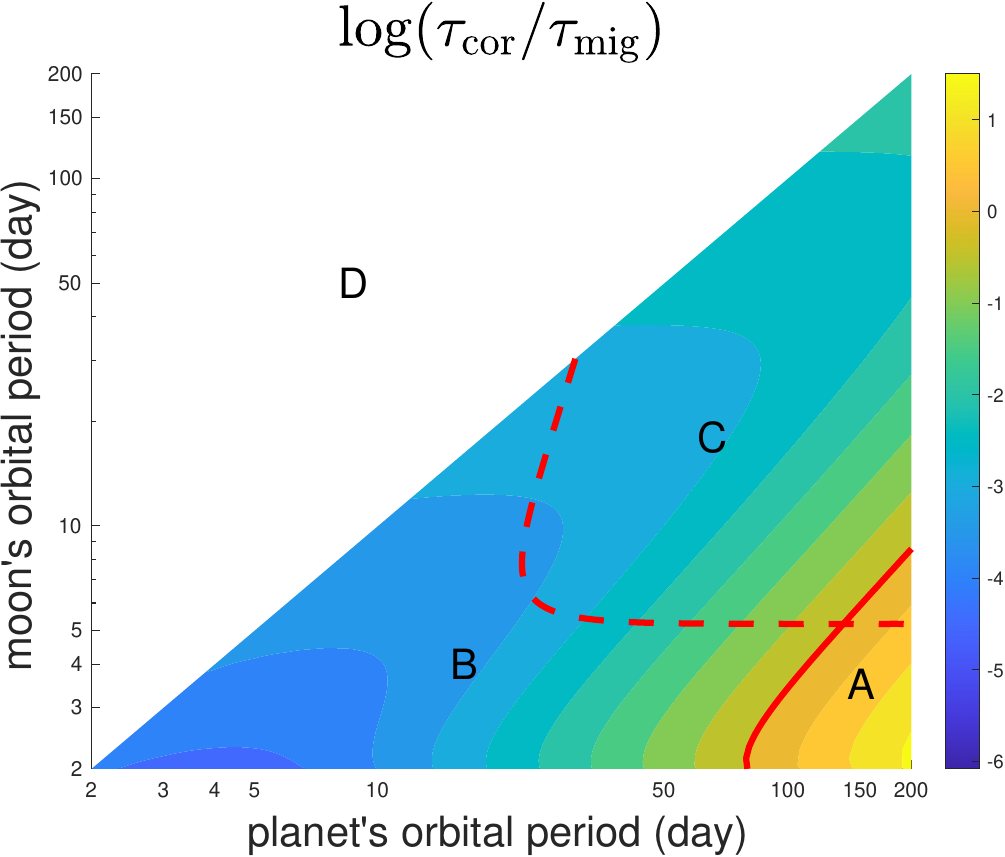}
\includegraphics[scale=0.5]{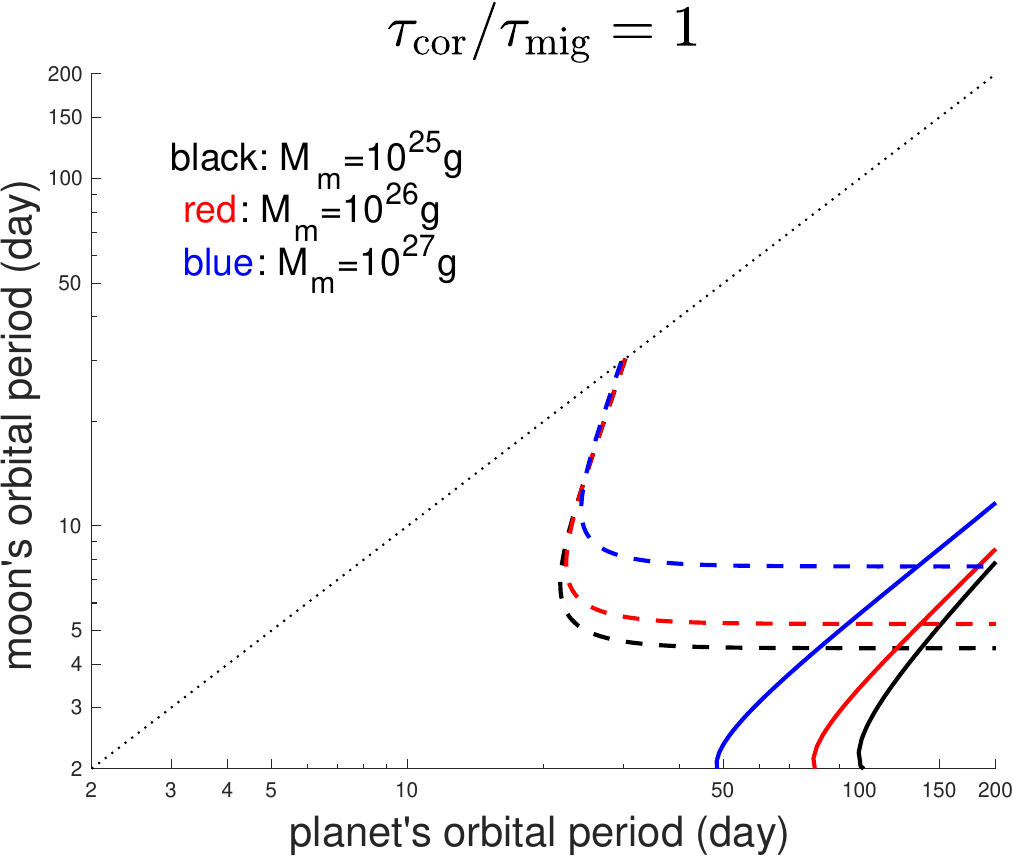}
\caption{Left panel: contours for $\log(\tau_{\rm cor}/\tau_{\rm mig})$ (Eqs. \ref{tau1} and \ref{eq:taumig}). Moon's and
planet's mass are $M_{\rm m} = 10^{26}$ g and $M_{\rm p} = M_{\rm J}$ respectively (Table \ref{table2}). Domain A denotes the 
ratio $\tau_{\rm cor}/\tau_{\rm mig}>1$, favourable 
for moon's survival. Domain B denotes $\tau_{\rm cor}/\tau_{\rm mig}<1$ with short $\tau_{\rm mig}<10~{\rm Gyr}$, unfavourable for moon's survival. Domain C 
denotes $\tau_{\rm cor}/\tau_{\rm mig}<1$ with long $\tau_{\rm mig}>10~{\rm Gyr}$ (longer than the age of planetary system), favorable for 
moon's survival. In order for a moon to be bound to the planet, $\omega_p<\omega_m$ such that top-left half domain D is a zone of avoidance.
Right panel: Investigation of moon's mass $M_{\rm m}=10^{25}$ g (black), $10^{26}$ g (red), and $10^{27}$ g (blue). 
In all cases $M_{\rm p} = M_{\rm J}$ and density of the moon is taken to be
$3~\rm g/cm^3$.
Solid curves denote ratio $\tau_{\rm cor}/\tau_{\rm mig}=1$, on the right of which moon can retain. Above dahsed lines $\tau_{\rm mig}>10$ Gyr and moon can retain.} \label{fig:tau}
\end{figure*}

Since $\tau_{\rm cor}\propto a_p^6$ and $\tau_{\rm mig}\propto a_m^{6.5}$ (roughly), large $a_p$ and small $a_m$ favour the retention of exomoon, and Figure \ref{fig:tau} verifies this survival preference. The left panel shows the contours of ratio $\tau_{\rm cor}/\tau_{\rm mig}$ versus planet's and moon's orbital periods. 
At the end of stage \textcircled{2}, the moon is assumed to be outside corotation radius of its rapidly spinning 
host planet.  In domain A with $\tau_{\rm cor}/\tau_{\rm mig}>1$, the moon's orbit expands prior to the planet's spin down.  This region
is favorable for the moon's retention.   In domain B $\tau_{\rm mig}<\tau_\star \sim 10~{\rm Gyr}$ and in domain C 
$\tau_{\rm mig}>\tau_\star$.  Since $\tau_{\rm cor} < \tau_{\rm mig}$ in both domains B and C, they contain moons which 
undergo orbital decay after $r_{\rm cor}$ is reduced interior to $a_m$.  Moreover, moons would plunge into their host planet 
before  their host star evolves off the main sequence in domain B whereas they would be retained in domain C. Thus, \textit{\textbf{exomoons tend to be retained around cold but not hot Jupiters, unless the moon's migration timescale is 
longer than the age of planetary system}}. 
Although tidal torque $\Gamma_{pm}^t\propto M_m^2$, magnetic torques $\Gamma_{pm}^m$ and $\Gamma^{m}_{\star m} \propto M_m$ (magnetic torques $\propto R_m^3$ and moon's density is roughly 3 $\rm g/cm^3$). The moon's 
orbital angular momentum $\propto M_m$ so that the two migration timescales have different scaling laws with respect to moon's mass, i.e., tidal migration timescale $\propto M_m^{-1}$ but magnetic migration timescale is independent of $M_m$, and therefore a moon with larger mass corresponds to smaller migration timescale. The right panel
of Figure \ref{fig:tau} shows moon's retention for different moon's masses. A moon with larger mass is more likely to be found around a planet with shorter distance from its host star.

Finally, in order for moons to be bound to their host planets, they must be within the planetary Hill radius, 
{\it i.e.} $\omega_p<\omega_m$, such that the top-left area separated by the dotted line is a zone of avoidance. 
If the planets' spin frequency $\Omega_p$ is synchronized with their orbital frequency $\omega_p$ at the end of
stage \textcircled{3}, their moons' $a_m < r_{\rm H} = r_{\rm cor}$.  Planets' subsequent inward migration would further reduce their 
moons' survivable probability.  Although some host planets may continue to spin down with outward migration, $a_m < r_{\rm cor}$
(Fig. \ref{sketch2}) and their moons' orbit would not expand during stage \textcircled{4}.

\subsection{Numerical calculation of spin-orbit evolution}
\label{sec:numericalmodels}
We have estimated the timescales of planet's spindown and moon's migration. To investigate more rigorously we solve numerically planet's orbital and spin equations coupled with its moon's orbital equation in the circular and coplanar
limit.  A more general treatment is presented in 
Appendix \ref{sec:appendixnumericalmodels}.

Based on the above consideration, the evolution equations for $a_p$, $\Omega_p$ and $a_m$ become
\begin{equation}\label{dynamics}
\left\{
\begin{aligned}
\frac{d}{dt}\left(M_p(GM_\star a_p)^{1/2}\right)&=-\Gamma_{p\star}^t-\Gamma_{\star p}^m, \\
\frac{d}{dt}\left(\alpha_pM_pR_p^2\Omega_p\right)&=\Gamma_{p\star}^t, \\
\frac{d}{dt}\left(M_m(GM_pa_m)^{1/2}\right)&=-\Gamma_{pm}^t-\Gamma_{pm}^m-\Gamma^{m}_{\star m}.
\end{aligned}
\right.
\end{equation}
In the calculation of the magnetic torque, we adopt the evolution of planetary field $B_p\propto t^{-0.267}$ 
(Eq. \ref{evolution-b}) and 
assume the plasma density $N\propto t^{-2}a_p^{-2}$ (Eq. \ref{density}).
The numerical integration of the secular equations \eqref{dynamics} starts at the initial time when the planet has already formed through any mechanism: in-situ core accretion \citep{Bodenheimer2000} for a cold Jupiter, or high-e migration \citep{Rasio1996, Wu2003} or disk migration \citep{lin1996} for a hot Jupiter. This initial time is estimated at 2 Myr (see Table \ref{table1}). The initial $a_p$ is set such that the initial $2\pi/\omega_p$ is, respectively, 4 days, 10 days and 100 days. The initial $\Omega_p$ is set to be 40 hours. The initial $a_m$ is set to be $6R_J$ which is slightly outside corotation radius $r_{\rm cor}$ or $10R_J$ which is far outside $r_{\rm cor}$. The other parameters are listed in Table \ref{table2}.

\begin{figure*}
\centering
\includegraphics[scale=0.5]{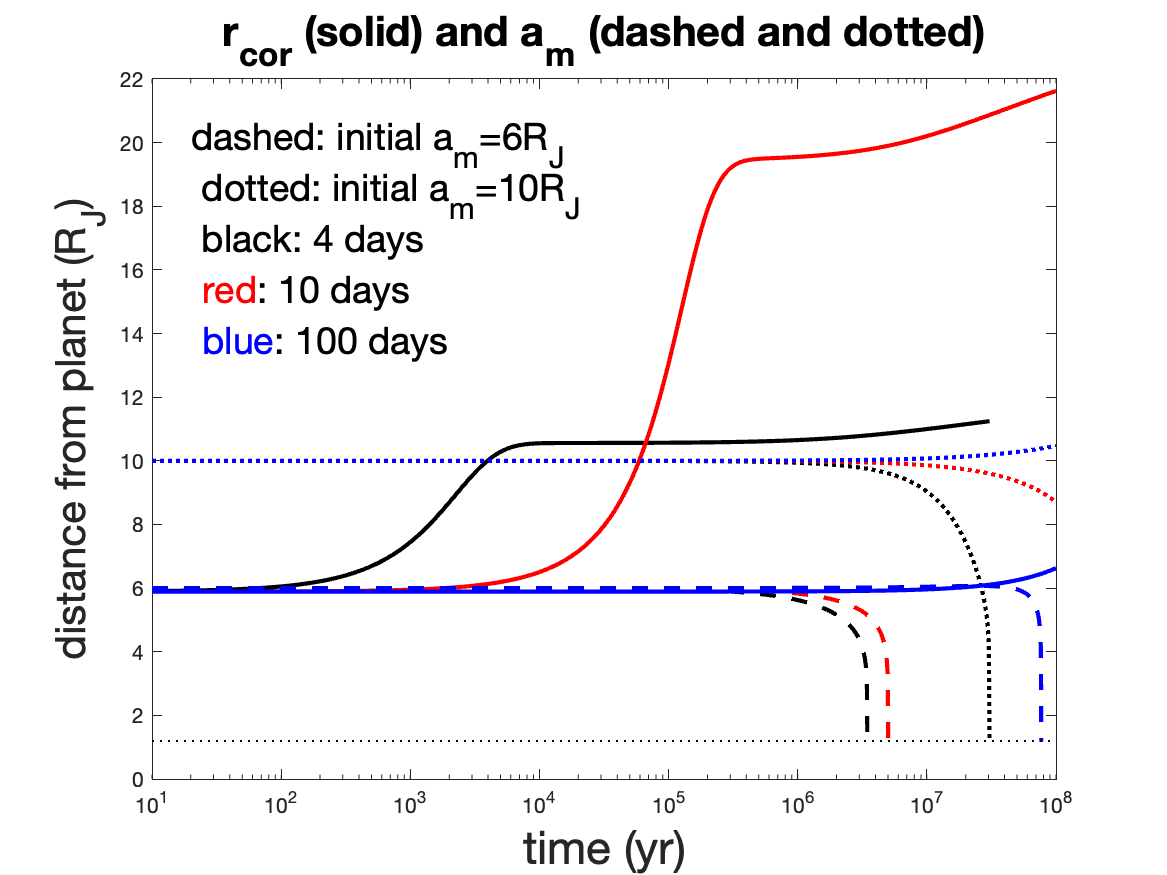}
\caption{The evolution of planet's corotation radius $r_{\rm cor}$ (solid) and moon's semi-major axis $a_m$ (dashed and dotted). Dashed lines: moon's initial orbit $a_m=6R_J$ slightly outside $r_{\rm cor}$, dotted lines: moon's initial orbit $a_m=10R_J$ a little far outside $r_{\rm cor}$. The initial planet's orbital period at 4 days (black), 10 days (red) and 100 days (blue). Horizontal thin dotted line denotes moon's Roche limit.}\label{dyn}
\end{figure*}

Figure \ref{dyn} shows the evolution of $r_{\rm cor}$ (solid) and $a_m$ (dashed and dotted) with different initial $a_p$ and $a_m$.
We firstly focus on the dashed lines with initial $a_m=6R_J$, sightly outside $r_{\rm cor}$. These values of $a_m$ and $M_m$ 
are comparable to those of Io. When a planet is initially close to its host star with an orbital period 
$2\pi/\omega_p=4$ days, the tidal torque $\Gamma_{p\star}^t$ raised by its host star is so strong that 
planet spins down very quickly. As the moon's orbit is engulfed by the expanding $r_{\rm cor}$ within 100 years, it migrates inward until it 
overflows its Roche lobe (horizontal think dotted line). With an initial $2 \pi/\omega_p=10$ days, moon's $r_{\rm cor}$ crosses
its $a_m$ within 2000 years. For a distant planet with $2 \pi/\omega_p=100$ days, planet's spin down is so slow 
that moon's $r_{\rm cor}$ remains inside $a_m$ until around $10^8$ years. At Jupiter's 
semi-major axis, $a_p \simeq 5.2$ au, the planet's spin retains its value at the end of stage \textcircled{2}. 
Since $a_m > r_{\rm cor}$, the moon migrates outward slightly. We then move to the dotted lines with initial $a_m=10R_J$ which is far outside $r_{\rm T}$ (Eq. \ref{eq:rtrun}). Moon with initial $2 \pi/\omega_p=4$ days has an initial $a_m > r_{\rm cor}$.  After 4000 years, $r_{\rm cor}$ expands
outside $a_m$ and the moon migrated inwards  until it overflows its Roche radius after $3\times10^7$ years. 
With the other two initial $2 \pi/ \omega_p$'s (10 and 100 days), moon is retained for at least $10^8$ years. 
Clearly, larger distance between moon and corotation radius prolongs the retention of the moon. In summary, the 
numerical calculations of the secular equations \eqref{dynamics} are consistent with the estimations of timescales 
in \S\ref{sec:migration}.

\section{Radio emissions}\label{sec:radio}

The star-planet or planet-moon magnetic interaction induces the radio emissions 
that can be detected by the radio telescopes, e.g. LOFAR, VLT, FAST, etc. 
\citep{zarka2007, zarka2019, vidotto2017, vidotto2022}. Recently, 
the detection of radio emission from the YZ Ceti system has been reported by
\citet{Pineda2023}. Since it contains several previously known super Earth\citep{AstudilloDefru2017}, these signals may 
be associated with the star-planet magnetic interaction \citep{Trigilio2023}.

The power of radio emissions due to the dipole-dipole or unipolar interaction can be estimated by the radio-magnetic Bode's law \citep{zarka2007}
\begin{equation}\label{power}
W_{\rm \star p, p m} \approx \eta_{\rm \star p, p m} v_{\rm \star p, p m} B_{\rm \star p, pm}^2 R_{\rm \star p, pm}^2 \pi/\mu
\end{equation}
where $v_{\rm \star p, p m} = (n_{\rm p, m} - \Omega_{\rm \star, p})a_{\rm p, m}$ is the relative velocity, 
$n_{\rm p, m}$ is the orbital frequency of the planet for the star-planet interaction or of the moon for the planet-moon interaction, 
$\Omega_{\rm \star, p}$ is the host's rotational frequency, and $a_{\rm p, m}$ is the orbital semi-major axis, 
$B_{\rm \star p, pm} = B_{\rm \star, p} (R_{\rm \star, p}/a_{\rm p, m})^3$ is the local magnetic field of the 
star/planet at the location of the planet/moon, 
%either $B=B_\star(R_\star/a_p)^3 = B_{\rm \star, p} (R_{\rm \star, p}/a_{\rm p, m})^3$ for the 
%star-planet interaction or $B=B_p(R_p/a_m)^3$ for the planet-moon interaction. 
$R_{\rm \star p, pm}$ is the obstacle radius with $R_{\rm \star p}\approx R_p (B_p/B_\star)^{1/3}(a_p/R_\star)$ 
for the star-planet interaction or $R_{\rm p m }\approx R_m$ for the planet-moon interaction. 
The dimensionless efficiency coefficient $\eta$ is about $2\times 10^{-3}$ according to the observations in the Jovian system
\citep{zarka2007}. We take the typical values of a young stellar system to estimate 
the radio power of the star-planet and planet-moon interactions. As before, the parameters we use 
are listed in Table \ref{table2} for a star-planet-moon system. The 
estimations show that the radio power of star-planet magnetic 
interaction is $2.1\times 10^{26}$ erg/s and that of planet-moon magnetic interaction is $2.3\times 10^{22}$ erg/s. 
For more mature systems with much weaker $B_\star$ and $B_{\rm p}$ (Table \ref{table2}, these powers
are much reduced.

The averaged flux density (power per unit area per unit frequency) is \citep{vidotto2022}
\begin{equation}\label{flux}
F_{\rm \star p, pm}=W_{\rm \star p, pm}/(\Theta d^2\Delta\nu)
\end{equation}
where $d$ is the distance of the planetary system from the Earth, $\Theta$ is the solid angle of the radio emission 
(1.58 sr for Jupiter-Io interaction), $\Delta\nu$ is the waveband of the radio telescope. We use this formula to 
estimate the average flux density for different sources and different telescopes. The energy spectrum of cyclotron 
radiation peaks at Larmor frequency and its double frequency. 

Suppose that the radio telescope wavebands to detect the 
star-planet and planet-moon unipolar interactions focus near $\sim$ 100 MHz (LOFAR at 10 $\sim$ 240 MHz and the 
future FAST at 70 $\sim$ 200 MHz), and that the distance from the earth is 1 kpc, then the 
averaged flux density of the star-planet interaction is $F\approx 14$ mJy and that of the planet-moon 
interaction is $F\approx 1.5\times 10^{-3}$ mJy. We can then estimate the upper limit on detectable distance 
for the radio telescopes. For example, the detection sensitivity of FAST is about 1 mJy and it can detect 
radio signals from star-planet magnetic interaction out to $\sim$ 4 kpc, and the planet-moon radio 
signals can be detected within $\sim$ 40 pc. It is possible to detect star-planet interaction
well beyond the distances ($\sim 10^2$ pc) of the closest Scorpius-Centaurus OB association\citep{blaauw1964ARA&A} 
and the well-studied Taurus and Orion star formation regions.  But radio emission from planet-moons magnetic 
interaction may be marginally detectable, at best, for only super-Jupiter young ($\lesssim 10$Myr) 
planets (Table \ref{table1}).

What we considered in the above is a young star-planet or planet-moon system. For a mature star-planet or 
planet-moon system, the stellar field drops from $\sim$ 1000 gauss to $\sim$ 1 gauss and the planetary field 
from $\sim$ 100 gauss to $\sim$ 10 gauss, and the upper limit of detection distance which is almost proportional 
to magnetic field (Eqs. \ref{power} and \ref{flux}) will be $\sim$ 13 pc for a mature star-planet system or 
$\sim$ 4 pc for a mature planet-moon system. For a mature star-planet-moon system, the radio emission from the 
planet-moon interaction is comparable to that from the star-planet interaction.

\section{Summary and Discussions}\label{sec:discussions}

In this paper we derive the scaling law for planet magnetic field through MHD dynamo equations. The scaling law depends on mass, radius and
luminosity but it is insensitive to rotation. Then we combine this scaling law with the virial theorem and the planetary evolution track to find the evolution law of magnetic field of the young Jupiter and of the exoplanets with different mass. Next we derive the magnetic torques induced by dipole-dipole and unipolar interactions for star-planet, planet-moon and star-moon systems. As proposed in \citet{lin1996}, a strong field in stellar magnetosphere can terminate the migration of short-period exoplanet, and similarly, such a strong field of young Jupiter can terminate the migration of Io at its present location.

We then investigate the possibility to find exomoons. By comparing the inner and outer radii of circumplanetary disk, we find that too strong planetary fields or too short star-planet distance do not favor the retention of exomoons. By comparing the planet's spindown and moon's migration timescales and numerically computing the 
equations of dynamics, we find that large exomoons are vulnerable to merge 
with their mature hot-Jupiter planets because the planets' spin and 
corotation radius evolve on shorter timescale than that of the exomoon's 
orbital migration. Nevertheless, the debris of tidally disrupted 
exomoons can lead to planetary rings which may be potentially detectable.
Observational inferences of such rings can provide supportive evidences
for disk over high-e migration scenario for hot Jupiters' origin   
since dynamical instabilities during the highly-eccentric close stellar 
encounters
are likely to effectively dislodge most of the exomoons from the 
gravitational confinement of their host planets.  

Provided they have similar densities, the magnetic torques 
$\Gamma^m _{p m}$ and $\Gamma^m _{\star m}$ increases with moons' 
mass. Moreover, due to their much weaker tidal torque $\Gamma 
^t _{pm}$, small exomoons (sub-Moon) have longer migration timescale, especially at relatively large distances ($\gtrsim 10 R_J$) 
from their less massive (sub-Jupiter) host planets (with relatively 
weak fields, Eqs. \ref{evolution-b}, \ref{moon-magnetic-torque}, 
\ref{eq:tidallag},  \& \ref{eq:taumig}). Although 
they could survive around hot Jupiters if their $\tau_m$ is longer 
than the age of the planetary systems, small moons are much less 
observationally conspicuous.  In contrast, the survivability of 
potentially detectable (i.e. relatively large, super-lunar-size) 
exomoons are much higher around warm or cold Jupiters because 
their spins do not have time to become tidally synchronized 
with their orbits around their host stars. But the transit-detection 
probability also decreases with the host planets decrease with their 
semi-major axis. 

Finally, we estimate the power and flux density 
of radio emissions due to the star-planet and planet-moon magnetic 
interactions, and estimate the upper limit of distance to be detectable 
by FAST.

This study is primarily based on a simple prescription for the planets' 
magnetic field during the evolution of planetary radius and luminosity 
(\S\ref{sec:evolution}).  In general, magnetic field may also influences 
the planetary evolution to some extent. The interaction of zonal wind 
(or thermal current) in planetary atmosphere and planetary poloidal 
field can induce a strong electric current penetrating into the planetary 
interior. The ohmic dissipation associated with this electric current may be sufficiently strong to lead to the planetary inflation to compensate the gravitational contraction. This mechanism has been suggested to be an explanation for the anomalous large radii of hot Jupiters, and it was firstly studied in a kinematic regime with an externally prescribed field \citep{batygin2010} and then in a dynamical regime in which the Lorentz force is considered to be a drag on fluid motion \citep{batygin2011}. This hypothesis has been challenged by follow-up studies \citep{huang2012, wu2013} and numerical simulations \citep{rogers2014}. However, a Bayesian analysis of observational data \citep{fortney2018} shows that ohmic dissipation can be attributed as the culprit of the inflation of hot Jupiters with high mass. In the present context, the coupling between the magnetic field and the thermal current and its effect on the planetary inflation are not directly related to dynamo that we study in this paper. Dynamo is a self-excited field but not a prescribed one, and its energy source is the convective heat flux in the interior. Ohmic dissipation releases heat but it is negligible compared to the convective heat flux. From the point view of the second law of thermodynamics, the convective heat flux can do work to cause fluid motion and generate magnetic field, but ohmic dissipation only releases useless heat that cannot do work on fluid to generate magnetic field. Therefore, the effect of magnetic field on the surface flow and thermal current is unlikely to have a significant impact on the evolution of planetary magnetic field.

On the other hand, the planetary inner core may have an impact on dynamo because the existence of a large core alters the thermal structure of a planet. At the young age of our solar system, the collisions may be frequent due to the high eccentricity of planetesimals arising from the sweeping secular resonance \citep{zheng2017a,zheng2017b}, such that a relatively large dilute core can form in Jupiter \citep{liu2019}. The effect of a large core on the evolution of planetary magnetic field needs to be further studied.

With respect to application to the orbital dynamics, we consider only the circumstance of a co-planar circular orbit. The more complicated orbital dynamics about the semi-major axis, eccentricity and inclination evolution coupled with the magnetic field evolution will be further studied (see Appendix \ref{sec:appendixnumericalmodels}).

\section*{Acknowledgements}
We thank Andrew Cumming for his valuable contribution, extensive discussion, and insightful suggestions. We also thank Gary Glatzmaier, Fabio deColle, Gongjie Li and Chen Chen for useful discussions and Xueshan Zhao for drawing some schematic pictures.

\appendix
\section{Scaling law of planetary magnetic field}

We start from the total energy equation
\begin{equation}\label{energy1}
\partial_t(\rho v^2/2+B^2/2\mu)=-\bm\nabla\cdot\bm A+\delta\rho\,\bm g\cdot\bm v-D_\nu-J^2/\sigma.
\end{equation}
On the RHS, the vector $\bm A$ is the total energy flux. The second term is the buoyancy power for dynamo arising from gravitational contraction on the Kelvin-Helmholtz timescale (by comparison the dynamo timescale is much shorter than K-H timescale). The third term $D_\nu$ is viscous dissipation, and the last term is ohmic dissipation, where $\sigma=1/(\mu\eta)$ is electric conductivity and $\mu$ magnetic permeability. When magnetic field grows sufficiently strong, the back reaction of Lorentz force on flow results in the dynamo saturation, at which the total energy is statistically steady \citep{christensen2010, jones2014} and the LHS vanishes. We integrate \eqref{energy1} over the convective zone of dynamo, the net flux $\oint\bm A\cdot d\bm S$ is negligible compared to the energy in the interior and the viscous dissipation $D_\nu$ is negligible compared to ohmic dissipation. Eventually, we arrive at $\langle\delta\rho\,gv\rangle\approx\langle J^2/\sigma\rangle$ (brackets denote the volume average) which portrays the balance between the power and the dissipation for dynamo in a statistically steady state. 

Next we introduce the two length scales, the mixing length $l$ for turbulent momentum transport, i.e. the scale of largest turbulent eddies, and the small scale $l_B$ of magnetic field on which magnetic diffusion takes place. In stellar convection zone the mixing length is approximately twice of pressure scale height $l\approx 2p/(dp/dr)\approx 2p/(\rho g)\approx 2\mathcal{R}T/(\mu_m g)\approx c_pT/g$ where hydrostatic balance, equation of state for ideal gas ($\mathcal{R}$ is gas constant and $\mu_m$ is mean molecular weight) and $c_p=2.5\mathcal{R}/\mu_m$ in convection zone are employed. The stellar mixing length is roughly 1/10 of stellar radius. However, in planetary interior the mixing length cannot exceed the depth of convection zone due to the small size of planet. To estimate the length scale $l_B$ of magnetic field we use the balance between magnetic induction and magnetic diffusion, i.e., the former takes place on planetary radius $R$ while the latter on $l_B$, and readily obtain $l_B/l\approx(\eta/vl)^{1/2}=Rm^{-1/2}$ where magnetic Reynolds number $Rm=vl/\eta$ is defined on the mixing length $l$. This two-scale analysis is necessary for such a dissipative system (another well-known example is the boundary-layer analysis in fluid mechanics). Otherwise, if we admitted that the magnetic induction and the magnetic diffusion take place on the same scale then we would find that the magnetic Reynolds number is of order of unity at which dynamo cannot be driven. Inserting Ampere's law $J\approx B/(\mu l_B)$ with the estimation $l_B\approx(\eta l/v)^{1/2}$ into the balance $\langle\delta\rho\,gv\rangle\approx\langle J^2/\sigma\rangle$, we arrive at $\langle B^2/\mu\rangle\approx\langle\delta\rho\,gl\rangle$,
which indicates the equipartition between magnetic energy and buoyancy energy. In the mixing length theory, buoyancy energy is comparable to kinetic energy, i.e. $\delta\rho\,gl\approx\rho v^2$, and thus we obtain $\langle B^2/\mu\rangle\approx\langle\rho v^2\rangle$. 

To estimate turbulent velocity $v$ we introduce heat flux $F=\rho c_p\,\delta T\,v=c_pT\delta\rho\,v\approx c_pT\rho v^3/(gl)$ where $\delta T/T=-\delta\rho/\rho$ for ideal gas at constant pressure and the mixing length theory $\delta\rho\,gl\approx\rho v^2$ are employed, and then we obtain the estimation of turbulent velocity $v\approx\left[Fgl/(\rho c_p T)\right]^{1/3}$. In stellar convection zone the mixing length $l\approx c_pT/g$ such that $v\approx(F/\rho)^{1/3}$. This estimation about turbulent velocity is numerically validated by \citet{chan1996} and \citet{cai2014}. In solar convection zone the overall turbulent velocity is estimated around 30 m/s which is consistent with the asteroseismological observation \citep{hanasoge2012}. Inserting the estimation of $v$ into the balance $\langle B^2/\mu\rangle\approx\langle\rho v^2\rangle$ we obtain the estimation of magnetic energy $\langle B^2/\mu\rangle\approx\langle\rho^{1/3}F^{2/3}\rangle$. However, in planetary interior, due to the small size of planet we replace $l$ with $\min(l,d)$, where $d$ is the depth of planetary convection zone, such that the estimation of magnetic energy becomes $\langle B^2/\mu\rangle\approx\left\langle\rho^{1/3}F^{2/3}[\min(l,d))/(c_pT/g)]^{2/3}\right\rangle$ where the factor $\min(l,d)/(c_pT/g)$ is dimensionless. Clearly, this scaling law is independent of rotation since Coriolis force does not enter the energy equation that we used to derive this scaling law. Although rotation is not the driving mechanism it plays an important role in field geometry \citep{christensen2010}. Usually faster rotation leads to more dipolar field on a larger scale due to the columnar structure of fluid flow. In some scaling laws, field strength depends on rotation, e.g. \citet{stevenson1979}, \citet{davidson2013} and \citet{wei2022}. In the anisotropic rotating turbulence, the mixing length theory, which states the balance between buoyancy force and inertial force, cannot hold but Coriolis force enters the force balance, such that the field strength will eventually depend on rotation \citep{wei2022}. This inference is valid for stellar convection zone and observations show that stellar fields indeed depend on rotation \citep{wright2011, vidotto2014}. Planet is much smaller than star and the mixing length is usually taken to be the depth of planetary convection zone. Therefore, the rotational effect on the mixing length theory is negligible for planetary dynamo.

Next we consider the planetary internal structure. We normalize the density $\rho$ with the mean density $\rho_m$ and the heat flux $F$ with the surface heat flux $F_s$ such that the estimation can be written as $\langle B^2/\mu\rangle\approx\beta\rho_m^{1/3}F_s^{2/3}$ where $\beta=\left\langle(\rho/\rho_m)^{1/3}(F/F_s)^{2/3}\left[\min(l,d)/(c_pT/g)\right]^{2/3}\right\rangle$ is a structure factor. We simply assume $F/F_s\approx(R/r)^2$ in gas giant planet. Moreover, for a gas giant planet, the polytropic model can well describe its internal structure with the polytropic index $n$ ranging from 1 of partial degeneracy at low density to 1.5 of complete degeneracy at high density \citep{lissauer2013}. Using Jupiter's parameters we solve the Lane-Emden equation to find that $\beta\approx 1.12$ for $n=1$ and $\beta\approx 0.85$ for $n=1.5$, so that $\beta\approx 1$. We rewrite $\langle B^2/\mu\rangle\approx\rho_m^{1/3}F_s^{2/3}$ with $\rho_m=3M/(4\pi R^3)$ where $M$ is planet mass and $F_s=L/(4\pi R^2)$ where $L$ is intrinsic luminosity (i.e., luminosity due to internal heat) then we are led to
\begin{equation}
\langle B^2/\mu\rangle \approx 0.115 M^{1/3}R^{-7/3}L^{2/3}.
\end{equation}

\section{Discussions on spin and orbital angular momentum in star-planet-moon system}
\label{sec:appendixnumericalmodels}

%{\color{red} What we have considered is a circular and co-planar orbit. In this section, we rigorously discuss the changes in eccentricity, inclination and spin due to the angular momentum transfer arising from tidal/magnetic torques and stellar wind in a star-planet-moon system. This part can be extended in the future studies.}

In this section, we systematically construct quantitative relations between several 
processes which can potentially lead to star's, planet's, and moon's 
orbital and spin evolution (in \S\ref{sec:starplanetmoontransfer}).  
In order to highlight the dominant mechanisms, some of these effects 
(such as planetary spin-down due to mass loss, tidally induced 
eccentricity excitation, and secular interaction) are neglected in 
this paper (\S\ref{sec:despin},
\S\ref{sec:tidaleccentricityexcitation},
\& \S\ref{sec:secular}). With these approximations, 
we derive the moon's spin (\S\ref{sec:semievol}) and
compute its orbital evolution (\S\ref{sec:numericalmodels}).

\subsection{Transfer in the star-planet-moon systems.}
\label{sec:starplanetmoontransfer}
Total angular momentum for star-planet-moon systems is 
\begin{equation}
J^{tot} _{\star p m} 
= L _{\star p} + L _{p m} + S _\star + S _p + S _m \ \ \ \ \ \ {\rm where}
\end{equation}
\begin{equation}
L _{\star p, p m} = M_{p, m} \omega_{p, m} 
a_{p, m}^2 (1-e_{p, m}^2)^{1/2} 
\label{eq:orbitalangmom}
\end{equation}
are the angular momentum of the star-planet's and planet-moon's orbits, 
\begin{equation}
S _{\star, p, m} = \alpha_{\star, p, m} M_{\star, p, m} R_{\star, p, m}^2 \Omega_{\star, p, m}
\label{eq:spinangmom}
\end{equation}
and $\alpha_{\star, p, m}$ are, separated by comma, the star's, planet's, 
and moon's spin angular momentum and coefficient of moment of inertia respectively. 
%We use the symbols $L$
%and $S$ to separately represent the orbital and spin angular momentum.  
The subscripts with commas separate,
in corresponding orders, components for star ($\star$), planet ($p$), moon ($m$), star-planet ($\star p$) and 
planet-moon ($p m$) systems respectively.

Stellar wind  ${\dot S}^w _\star$ and planetary evaporation ${\dot S}^w _p$ carried by their mass losses 
(${\dot M}^w _\star$ and ${\dot M}^w _p$) lead to net losses of $J^{tot} _{\star p m}$. Stars' and planets' 
contraction (${\dot R}_\star$ and ${\dot R}_p$) changes their $\Omega_\star$ and $\Omega_p$ but not their 
$S _\star$ and $S _p$.  For infant planets/moons, the effect of their migration through the protostellar /
circumeplanetary disks can also be included in ${\dot S}^w _p$ and ${\dot S}^w _m$.   Due to the 
combination of stellar/planetary wind and tidal/magnetic torque, the spin angular momenta evolve at rates, 
\begin{equation}
%    {\dot S}_\star = \Gamma ^{t,m} _{\star p} - {\dot S}^w _\star, \ \  {\dot S}_p = \Gamma ^{t, m} _{p \star} +\Gamma^{t,m}_{pm} - {\dot %S}^w _p,
%    \ \  {\dot S}_m = \Gamma^{t, m}_{m star} + \Gamma^{t, m}_{m p},
    {\dot S}_{\star, p, m} = \Gamma ^{t, m} _{\star p, p \star, m \star} +\Gamma^{t,m}_{\star m, p m, m p} - {\dot S}^w _{\star, p, m}. 
\label{eq:spinwind}
\end{equation}
Due to the combination of secular $(\Gamma^{sec} _{\star p m}$), tidal, and magnetic torque, the orbital angular momentum
of the star-planet and planet-moon systems evolve at rates
\begin{equation}
%    {\dot L}_{\star p}= \Gamma ^{sec} _{\star p m} - \Gamma^{t,m}_{\star p} - \Gamma^{t,m}_{p \star} \ \ \ \ {\rm and} \ \ \ \ 
%    {\dot L}_{p m} = - \Gamma ^{sec} _{\star p m} - \Gamma^{t,m}_{p m} - \Gamma^{t,m}_{m p}. 
    {\dot L}_{\star p, p m} = \pm \Gamma ^{sec} _{\star p m} - \Gamma^{t,m}_{\star p, p m} - \Gamma^{t,m}_{p \star, m p} 
\label{eq:ldotstarp}
\end{equation}
where the plus and minus signs for $\Gamma^{sec}_{\star p m} $
refer to the secular transfer of orbital angular momentum between the 
star-planet and planet-moon systems respectively.
With the neglect of the second-order terms $\Gamma^{t,m}_{\star m}$ and $\Gamma^{t,m}_{m\star}$, the total angular momentum budget of the system changes at a rate
\begin{equation}
    {\dot J}^{tot} _{\star p m}= 
    {\dot L} _{\star p} + {\dot L} _{p m} + {\dot S} _\star + {\dot S} _p + {\dot S} _m \simeq - {\dot S}^w _\star - {\dot S}^w _p - {\dot S}^w _m. 
\end{equation}

Neglecting structure changes (i.e. assuming ${\dot \alpha}_\star={\dot 
\alpha}_p={\dot \alpha}_m=0)$, the spin angular momentum, mass, and radius 
changes of the stars, planets, and moons in Equation (\ref{eq:spinangmom})
lead to spin rate evolving at rates
\begin{equation}
    {{\dot \Omega}_{\star, p, m} \over \Omega_{\star, p, m} }= {{\dot S}_{\star, p, m} \over S_{\star, p, m}} -
    {\dot M_{\star, p, m} \over M_{\star, p, m}} - {2 {\dot R}_{\star, p, m} \over R_{\star, p, m}}. 
    \label{eq:spindot}
\end{equation}
Changes in the star-planet and planet-moon's orbit 
in Equation (\ref{eq:orbitalangmom}) also lead to
\begin{equation}
    {{\dot a}_{p, m} \over 2 a_{p, m} }- {e_{p, m}{\dot e}_{p, m} \over 1-e_{p, m}^2} 
    = {{\dot L}_{\star p, p m} \over L_{\star p, pm}} - {{\dot M}_{p, m} \over M_{p, m}}
    - {{\dot M}_{\star, p} \over 2 M_{\star, p}}. 
    \label{eq:orbitdot}
\end{equation}

\subsection{Spin and mass loss}
\label{sec:despin}
In Eq (\ref{eq:spindot}), the effect of angular momentum and mass loss from 
stars and planets is included in ${\dot S}^w _\star$ and ${\dot S}^w _p$ 
through ${\dot S}_\star$ and ${\dot S}_p$ (Eq. {\ref{eq:spinwind}),
${\dot M}_\star$ and ${\dot M}_p$.  After the initial T Tauri phase, host stars' 
mass and radius evolution (${\dot M}_\star/M_\star$ and ${\dot R}_\star/R_\star$)
is negligible.  But, the loss of angular momentum is strongly enhanced by 
magnetic braking \citep{mestel1968}. An empirical formula for the
observed rotational velocity $\langle V_r\sin i\rangle$ (where $i$ is 
the inclination between the spin axis and the line of sight) of G 
and K stars with age $\tau_\star$ suggests $\Omega_\star R_\star 
\simeq V_0 (\tau_0/\tau_\star)^{0.5}$ where $V_0 \simeq 4$km s$^{-1}$ and 
$\tau_0 \simeq 1$ Gyr \citep{skumanich1972}. The corresponding
spin-down timescale 
\begin{equation}  
S_\star / {\dot S}^w _\star \simeq 
\Omega_\star/{\dot \Omega}^w _\star \sim 
-2 \tau_0 (V_0/\Omega_\star R_\star)^2 \sim -2 \tau_\star.
\end{equation}
For young systems, this loss timescale may be considerable
shorter than the tidal circulation timescale \citep{dobbsdixon2004}.
But for mature stars with slow $\Omega_\star$ and vanishing oblateness, 
angular momentum loss carried by stellar wind becomes negligible.

Photo-evaporation of close-in planets can also lead to significant
fractional loss of super-Earths' atmosphere \citep{owen2013, owen2017, 
fulton2017}. However, this photo-evaporation process is unlikely to 
be effective for Jupiter-mass planets. Under some circumstances
(such as high-e migration), strong tidal \citep{gu2003, gu2004} or 
ohmic \citep{laine2008, laine2012, hou2022} dissipation of host stars' 
gravitational and magnetic perturbation may lead to intense heating, 
runaway inflation, and mass loses. The presence of planets' magnetic 
field may enhance the their angular moment loss. Moons engulfed in 
such out-flowing envelopes would endure hydrodynamic drag and undergo
orbital decay.  Such effects are beyond the scope of the present
investigation.  In this paper, we assume ${\dot S}^w _p = {\dot M}_p =0$.

\subsection{Eccentricity evolution}
\label{sec:tidaleccentricityexcitation}
In Eqs. (\ref{eq:spindot}) and (\ref{eq:orbitdot}), ${\dot a}_{p, m}$, ${\dot e}_{p, m}$, and 
${\dot \Omega}_{\star, p, m}$ are separately treated through ${\dot L}$'s and ${\dot S}$'s respectively.  
Tidal dissipation inside rapidly/slowly spinning stars can excite/damp 
their planets' eccentricity \citep{goldreich1966, dobbsdixon2004}.  Likewise tidal dissipation inside rapidly/slowly spinning 
planets can also excite/damp their moons' eccentricity.  
In systems where 1) all spin vectors of the host stars, planets, and moons are aligned with the planets' and moons' orbital angular momentum 
vectors and 2) $a_m < < r_H$ (so that the stars' secular perturbation on the moons' orbits can be neglected), the governing equations 
for the eccentricity evolution become
\begin{equation*}
    {\dot e}_p = g_{\star p} + g_{p \star} \ \ \ \ \ \ {\rm and} \ \ \ \ \ \ {\dot e}_m \simeq g_{p m} + g_{m p} \ \ \ \ \ \ {\rm where}
\end{equation*}
\begin{align*}
    g_{p \star, \star p} = & 27k_{p,\star}\tau_{p,\star}\omega_p e_p {M_{\star, p} \over M_{p, \star}}
    {R_{p, \star}^5 \over a_p^5}  \\ & \cdot\left( f_2(e_p) {11 \Omega_{p, \star} 
    \over 18 \omega_p} - f_1 (e_p) \right),  \\
    g_{m p, p m} = & 27k_{m,p}\tau_{m,p}\omega_m e_m {M_{p, m} \over M_{m, p}}
    {R_{m, p}^5 \over a_m^5}  \\ & \cdot\left( f_2(e_m) {11 \Omega_{m, p} 
    \over 18 \omega_m} - f_1 (e_m) \right), 
\end{align*}
\begin{equation}
    f_1 (e_{p, m}) = {1+15 e_{p, m}^2/4 + 15 e_{p, m}^4 /8 + 5 e_{p, m}^6 /64 \over (1-e_{p, m}^2)^{13/2}}, \nonumber
\end{equation}
\begin{equation}
    f_2 (e_{p, m}) = {1+3 e_{p, m}^2/2 + e_{p, m}^4 /8 \over (1-e_{p, m}^2)^5}. \nonumber
\end{equation}

These equations indicate that around rapidly spinning stars (with $\Omega_\star$ or $\Omega_p >> \omega_p$),
tidal torque can excite the planets' eccentricity.  Similarly
around rapidly spinning planets (with $\Omega_p$ or $\Omega_m >> \omega_m$), tidal torque can excite 
the moons' eccentricity.  However, stellar wind leads to the rapid spindown of young solar type 
stars (\S\ref{sec:despin}). For hot Jupiters, the dominance $\Gamma^t _{p\star}$ over $\Gamma^t _{\star p}$ 
also leads to synchronization between the planets' spin with their orbit, i.e. $\Omega_p \sim \omega_p$ 
(\S\ref{sec:magtide}). Thus, tidal interaction between stars and their planets usually damps their 
eccentricity \citep{dobbsdixon2004}.  Indeed, nearly all the hot Jupiters and most of the 
warm Jupiters have circular orbits or very small eccentricities.  
Moreover, since the moons' $a_m < r_H$, their $\omega_m$ is greater 
than their planets' $\omega_p$ and therefore $\Omega_p$ ($\Omega_p$ with respect to hot Jupiters).  Tidal interaction between planets and their
moons also usually damps the moons' eccentricity.  In this paper, we consider the limiting
cases $e_p < < 1$ and $e_m < < 1$ so that the second term on the left hand side of Eq (\ref{eq:orbitdot})
is negligible.

\subsection{Secular interaction}
\label{sec:secular}
In Eq (\ref{eq:orbitdot}), we also consider the dissipationless 
gravitation interaction between the stars, planets, and moons
(included in $\Gamma^{sec}_{\star p m}$ through ${\dot L}_{\star p, p m}$,
Eq. \ref{eq:ldotstarp}). Although the Laplace-Runge-Lanz vector 
(eccentricity, periapsis longitude and ascending node) of stars, 
planets, and moons generally modulate on a secular timescale, 
3-body (star-planet-moon) gravitational interaction does not 
lead to a net angular momentum and energy exchanges over 
multiple liberation or circulation cycles \citep{murray1999}.  Gravity
associated with their axisymmetric natal disks, general relativity, and rotationally-flatten 
oblateness of their host stars/planets also do not lead to any net changes in 
planet and moons' angular momentum, albeit they lead to precession of 
planet's and moon's ascending nodes and periapsis longitudes.

However, during the depletion of their natal disks or the spin down of their host 
stars and planets (due to ${\dot S}^w _\star$ and ${\dot S}^w _p$),
these systems may encounter secular resonances with net angular 
momentum (i.e. finite ${\dot L}^{sec} _{\star p m}/J^{tot} _{\star p m}$) 
and negligible energy (i.e. conserved $a_p$ and $a_m$) exchanges between the 
star-planet and planet-moon's orbits \citep{nagasawa2003, nagasawa2005}.
Moreover, when the planet's/moon's apsidal and nodal precession frequency 
match with their mean motion ($\omega_p$ and $\omega_m$), their 
eccentricity and inclination may also be excited through the evection and eviction 
resonances in coplanar and inclined systems respectively \citep{touma1998}.  
Together with the stellar, planetary, and moons' tidal torque, these physical 
processes may have played a role in the dynamical history of the 
Sun-Earth-Moon system \citep{zahnle2015, cuk2016}.  They are also likely 
to disrupt satellite systems formed in the already confined birth-domains 
(\S\ref{sec:radii}) around short-period planets which may have migrated to
their present-day location.  In this paper, we neglect these dynamical 
effects.

\subsection{The rate of change in the semi-major axis}
\label{sec:semievol}
In Eqs (\ref{eq:spindot}) and (\ref{eq:orbitdot}), we have listed all the contributing factors
which may affect the spin-orbit evolution of a star-planet-moon system, under the influence of
tidal, magnetic, and secular interactions as well as angular momentum and mass loss, and stellar
and planetary contraction. Some of these contributing factors, such as secular interaction,
stellar and planetary winds, are important during the system's formation and infancy epochs.
While these effects will be explored in subsequent studies, this paper focuses on the long-term
survival of exomoons on the orbital period of their host planets.  The above discussions 
indicate that most these effects have negligible contribution in mature (Gyr old) systems.

Based on the above discussions, we neglect, for mature star-planet-moon system, contributions from  
${\dot S}^{sec} _{\star p m}$, ${\dot S} ^w _{\star, p}$, ${\dot M}_{\star, p, m}$, 
${\dot R}_{\star, p, m}$, $e_{p, m}$, so that Eqs (\ref{eq:spindot}) and (\ref{eq:orbitdot}) reduce to 
%\begin{align*}
%   & 
\begin{equation}
   {{\dot \Omega}_{\star, p, m} \over \Omega_{\star, p, m} } = 
    {\Gamma ^{t, m} _{\star p, p \star, m \star} +\Gamma^{t,m}_{\star m, p m, m p}
    \over S_{\star, p, m}}, 
    {{\dot a}_{p} \over 2 a_{p} } =
    {- \Gamma^{t,m}_{\star p} - \Gamma^{t,m}_{p \star} \over L_{\star p}}, 
\end{equation}
\begin{equation}
%    \\
%    & 
    {{\dot a}_{m} \over 2 a_{m} } =
    {- \Gamma^{t,m}_{p m} - \Gamma^{t,m}_{m p} - \Gamma^{t,m}_{m\star} - \Gamma^{t,m}_{\star m} \over L_{p m}}.
    \end{equation}
%\end{align*}
We focus on the orbital evolution under the assumption that the eccentricities of the moon's 
orbit around the planet and the planet's orbit around the star are quickly damped by $\Gamma ^t _{mp}$ 
or $\Gamma^m _{mp}$ and $\Gamma ^t _{p \star}$ or $\Gamma^m _{p \star}$ respectively.
The effects introduced by the initial stellar and planetary spin with eccentricity and inclination in young systems will be
analyzed in a follow-up investigation.}

\bibliographystyle{apj}
\bibliography{paper}

\end{CJK*}
\end{document}